\documentclass[10pt]{IEEEtran}
\usepackage{setspace}
\usepackage{relsize}
\usepackage{xcolor}
\usepackage{combelow}
\usepackage{comment}
\usepackage{amsmath,amssymb,amsfonts}
\usepackage{algorithmic}
\usepackage{textcomp}
\def\BibTeX{{\rm B\kern-.05em{\sc i\kern-.025em b}\kern-.08em
    T\kern-.1667em\lower.7ex\hbox{E}\kern-.125emX}}
\usepackage{tikz,graphicx}
\usepackage{color}
\usepackage[mathscr]{eucal}

\def\XXint#1#2#3{{\setbox0=\hbox{$#1{#2#3}{\int}$}
     \vcenter{\hbox{$#2#3$}}\kern-.5\wd0}}

\newcommand{\be}[1]{\begin{equation} \label{#1} }
\newcommand{\bea}[1]{\begin{eqnarray} \label{#1} }

\newcommand{\bfi}{\begin{figure}}
\newcommand{\efi}{\end{figure}} 
\newcommand{\ee}{\end{equation}}
\newcommand{\eea}{\end{eqnarray}}
\newcommand{\bib}{\bibitem}
\newcommand{\lbl}{\label}

\newcommand{\w}{{\omega}}

\newcommand{\vrp}{{\bf r}}

\newcommand{\vrh}{{\bf \hat{r}}}

\newcommand{\vr}{{\bf r}}
\newcommand{\vH}{{\bf H}}
\newcommand{\vE}{{\bf E}}

\newcommand{\vB}{{\bf B}}
\newcommand{\vF}{{\bf F}}
\newcommand{\vD}{{\bf D}}

\newcommand{\dmu}{\overline{\mbox{\boldmath $\mu$}}}
\newcommand{\dep}{\overline{\mbox{\boldmath $\epsilon$}}}
\newcommand{\dnu}{\overline{\mbox{\boldmath $\nu$}}}
\newcommand{\dtau}{\overline{\mbox{\boldmath $\tau$}}}

\newcommand{\V}{{\mathcal V}}

\newcommand{\vthh}{{\bf \hat{\mbox{\boldmath {$\theta$}}}}}

\newcommand{\la}{\lesssim}
\newcommand{\ga}{\gtrsim}
\newcommand{\eps}{\epsilon}

\newcommand{\cQ}{{\cal Q}}
\newcommand{\cW}{{\cal W}}

\begin{document}
\title{\LARGE\bf Fundamentals of Antenna Bandwidth and Quality Factors}
\author{Arthur D. Yaghjian\\\textit{\normalsize Electromagnetics Research, Concord, MA  01742,  USA} ({\small a.yaghjian@comcast.net}) %, Second B. Author, and Third C. Author, Jr., \IEEEmembership{Member, IEEE}
%\thanks{Manuscript ... .  This work was supported in part under the U.S. Air Force Office of Scientific Research Contract \# FA9550-19-1-0097 through Dr. Arje Nachman.}
%\thanks{A.D. Yaghjian works as an Electromagnetics Research Consultant, Concord, MA 01742 USA (e-mail: a.yaghjian@comcast.net).}
}
\maketitle
\thispagestyle{empty}
{\bf \boldmath
\begin{abstract}
 After a brief history of the development of quality factors, useful expressions are derived for the robust input-impedance $Q_Z(\w)$ quality factor that accurately determines the (voltage-standing-wave-ratio) VSWR fractional bandwidth of antennas for isolated resonances and a small enough bandwidth power drop.  For closely spaced multiple resonances/antiresonances, a definitive formula is given for the increase in fractional bandwidth enabled by Bode-Fano tuning.  Methods are given for determining the conventional and complex-energy quality factors of antennas from RLC circuit models.  New field-based quality factors $Q(\w)$ are derived for antennas with known fields produced by an input current.  These $Q(\w)$ are remarkably robust because they equal $Q_Z(\w)$ when the input impedance is available.  Like $Q_Z(\w)$, the field-based $Q(\w)$ is independent of the choice of origin of the antenna fields and is impervious to extra lengths of transmission lines and surplus reactances.  These robust field-based quality factors are used to derive new lower bounds on the quality factors (upper bounds on the bandwidths) of spherical-mode antennas that improve upon the previous Chu/Collin-Rothschild (CR) lower bounds for spherical modes.  A criterion for antenna supergain is found by combining the Harrington maximum gain formula with the recently derived formula for the reactive power boundaries of antennas.  Maximum gain versus minimum quality factor for spherical antennas are determined using the improved lower bounds on quality factor for different values of electrical size $ka$.  Lastly, reduced antenna quality factors allowed by dispersive tuning overcome the traditional Chu/CR lower bounds for lower radiation efficiencies and small enough bandwidth power drops.
\end{abstract} 
\unboldmath
}
%
%\relscale{0.95}
%\setstretch{.9}
%
\section{Historical Introduction \lbl{Intro}}
\IEEEPARstart{T}{he} concept of quality factors in electrical engineering appears to have originated circa 1914 with the ``inductive purity'' $K\!\! =\! \!\w L/R$ defined at the angular frequency $\w$ by K.S. Johnson as the reciprocal of the ``dissipation factor'' for lossy coils with inductance $L$ and resistance $R$ \cite{Green,Smith}.  In 1918, the Circular 74 of the United States Bureau of Standards used the expression $\w L/R$ to define the ``sharpness of resonance''\footnote{When an initially open series RLC circuit with a charged capacitor is closed, the voltages and currents ``ring'' or ``resonate'' with damped oscillations (for $R$ not too large) at a ``natural'' frequency close to the tuned frequency $\w \!=\!\! 1/\sqrt{LC}$ of the RLC circuit; thus the term ``resonant frequency'' for $\w\!$ \cite{NBS-1918}.} of a series RLC circuit but without referring to it as quality factor or $Q$ \cite{NBS-1918}.  By 1920, Johnson had changed the ``overworked'' symbol $K$ to $Q$, which was not initially an abbreviation for ``quality'' or ``quality factor'', a term first used for Johnson's symbol $Q$ by his colleague V.E. Legg at Bell Telephone Laboratories (previously Western Electric) that became popular after about 1925.   In his 1932 book \cite{Terman-1932}, F.E. Terman, the ``father of Silicon Valley'', was apparently the first to note that the quality factor for a  series RLC resonance could be expressed as $Q_{\scriptscriptstyle\rm RLC}\!\! =\! \!\w W_{\scriptscriptstyle\rm LC}/P_{\scriptscriptstyle\rm R}$, the present-day IEEE general definition of quality factors \cite{IEEE}, where $W_{\scriptscriptstyle\rm LC}$ is the total average energy stored in the fields of the inductor plus capacitor and $P_{\scriptscriptstyle\rm R}$ is the average power dissipated in the resistor.  Moreover, Terman proved that the $-3$ dB  conductance fractional bandwidth of a series RLC circuit about its resonant frequency is equal to the inverse of the quality factor, namely $1/Q_{\scriptscriptstyle\rm RLC}$ [half the voltage-standing-wave-ratio (VSWR) fractional bandwidth]  \cite{Y&B-2005}.  Although not as universally applicable to antennas as VSWR fractional bandwidth, the conductance fractional bandwidth for series RLC circuits, or resistance bandwidth for parallel RLC circuits, has the advantage of not requiring the introduction of a single-mode feed waveguide with a terminal surface (port) and a characteristic impedance \cite{Y&B-2005, Kerns}.
\par
The present paper concentrates on the fundamentals of antenna bandwidth and quality factors with emphasis on the derivation and applicability of the VSWR impedance-bandwidth quality factor $Q_Z(\w)$, as well as on the generalized complex ``Q-energy'' from which a new improved field-based quality factor $Q(\w)$ is derived that has the same robust accuracy and applicability as $Q_Z(\w)$.  The recently discovered technique, based on the complex Q-energy, for the dispersive tuning of antennas is highlighted as a practical means to overcome the traditional Chu lower bounds on quality factors.  There have been relatively few significant advancements in the fundamentals of antenna bandwidth and quality factors since the seminal papers of Wheeler \cite{Wheeler-1947}, Chu \cite{Chu}, Collin-Rothschild \cite{C&R}, and Fante \cite{Fante-1969}, as opposed to the many papers that have been published on applying, evaluating, and optimizing the fundamental expressions for the bandwidths and quality factors of different classes and designs of antennas.  There have been a number of noteworthy review papers within this derivative area of antenna research, including review articles on the physical bounds and stored energies of antennas and other radiating systems \cite{G&T&C}, \cite{Schab}.  Nevertheless, the present paper confines itself to the more fundamental developments of antenna bandwidth and quality factors.
\subsection{RLC Circuit Approximation for the Impedance and Quality Factor of Antennas\lbl{IntroA}}
The quality factors for linear, passive, time invariant, transmitting\footnote{The bandwidth of a receiving antenna is limited only by the signal to noise ratio of the output amplifiers and the noise in the input signal to the antenna.  Several papers have been written on the receiving bandwidth of antennas, for example, \cite{Best-2016, Sarabandi&Rao,Loghmannia&Manteghi}, and thus receiving antennas are not treated in the present paper.  Also, non-Foster impedance matching and time-varying antennas, both of  which require active networks, are not considered herein.} antennas, fed by a voltage $V$ and current $I$ across a terminal-pair (or port of a single-mode waveguide) and tuned to an isolated resonance or antiresonance, can be found by simply approximating the input impedance (resistance $R$ and reactance $X$) of the antenna near the resonance ($dX/d\w>0$) or antiresonance ($dX/d\w<0$) with a series or parallel RLC circuit, respectively, provided the input impedance $Z(\w)\! =\! V/I\! =\! R(\w)\!+\! jX(\w)$ of the antenna is known as a function of frequency.  [Time-harmonic $e^{j\w t}$ ($\w>0$) dependence is used throughout.]  That is, the quality factor of the tuned antenna can be approximated with frequecy independent $R$,$L$, and $C$ such that $Q_{\scriptscriptstyle\rm RLC}\!\!=\!\!\w L/R\!=\!1/(\w CR)$ for a resonance and $Q_{\scriptscriptstyle\rm RLC}\!\!=\!\!R/(\w L)\!\! =\!\!\w CR$ for an antiresonance, respectively, where $L$ and $C$ can be found from the value of $dX(\w)/d\w$ for the tuned antenna along with the relationship $\w^2 LC \!=\!1$ ($\w \!> \!0$), and $R$ is the resistance at the resonant or antiresonant frequency.  Then the $-3$ dB conductance (or resistance) fractional bandwidth, which does not require a feed waveguide with a characteristic impedance for its definition (just a terminal-pair with $V$ and $I$), is given approximately as $1/Q_{\scriptscriptstyle\rm RLC}$ for a resonance (or antiresonance).\footnote{For $|dR(\w)/d\w|\!\!\ll \!\!|dX(\w)/d\w|$, the $\alpha$ power-drop fractional conductance bandwidth, $FCBW_\alpha(\w)\!=\! 2\Delta\w/\w$ (half-bandwidth $\Delta\w$), for a fixed input voltage applied to the antenna is determined approximately at the tuned resonant frequency $\w$ [$X(\w)\!=\!0, dX(\w)/d\w\! >\!0$] from the solutions to the equation \cite{Y&B-2005}\\[-3mm]
\be{cbw}
{\rm Re}[1/Z(\w \pm \Delta\w)] = \frac{R(\w)}{R^2(\w) + X^2(\w \pm \Delta\w)} = \frac{1-\alpha}{R(\w)}\\[-1mm]
\ee
to $O[(\Delta\w/\w)^2]$, where $FCBW_\alpha(\w) \approx \sqrt{\beta}/Q_{\scriptscriptstyle\rm RLC}$ with $\beta = \alpha/(1-\alpha)$.  The same inverse relationship with $Q_{\scriptscriptstyle\rm RLC}$ holds for the resistance bandwidth at an antiresonant frequency for a fixed input current.}
\par
An important difference between the RLC impedance approximation of a tuned antenna and a simple RLC circuit made of lumped ideal frequency independent circuit elements is that the frequency derivative $dR(\w)/d\w$ of the input resistance of an antenna can be nonzero even at a resonant or antiresonant frequency, and this can degrade the accuracy of the inverse relationship between the conventional RLC quality factor and the fractional bandwidth of an antenna, as explained below in Section \ref{CEQ}.  The utility of the quality factor of an antenna is rooted in its capability of accurately predicting the bandwidth of the antenna, and the primary importance of bandwidth lies in its proportionality relationship to the channel capacity.\footnote{In the context of antennas, the Shannon channel capacity \cite[th. 17]{Shannon-1948}\\[-3mm]
\be{sh}
\mathbb{C} = \mathbb{B} \log_2(1 + \mathbb{S}/\mathbb{N})\\[-2mm]
\ee
is the upper bound on the information rate in bits/s (with error $\!\!\to\!\! 0$) for a time-varying signal [for example, the voltage or current at the output port (or terminal-pair) of a receiving antenna] with a finite rectangular bandwidth $\mathbb{B}$ and an average available power $\mathbb{S}$, to which is added a total average noise power $\mathbb{N}$ that is both Gaussian and white (AWGN).  A ``channel'' does not come into play except as a possible source of the noise and a transporter of the signal from an input transmitting antenna port to the output receiving antenna port.  The ``channel'' is defined as everything between the input and output ports of the transmitting and receiving antennas, including the antennas.  The AWGN can enter at any point in the channel. (Because the bandwidth $\mathbb{B}$ is a strictly finite rectangular bandwidth, the signal cannot be strictly, just approximately, timelimited.)  Since white noise has a constant power spectral density $\mathbb{N}_0$ W/Hz,  $\mathbb{N}$ equals $\mathbb{N}_0\mathbb{B}$, where $\mathbb{N}_0$ can be expressed in terms of an \textit{effective} noise temperature $\mathbb{N}_0 = \mathcal{K}\mathcal{T}$ with $\mathcal{K}$ Boltzmann's constant.  As the bandwidth $\mathbb{B}$ approaches $\infty$ with $\mathbb{S}/\mathbb{N}_0$ fixed, the channel capacity $\mathbb{C}$ approaches the value $(\log_2 e)\,\mathbb{S}/\mathbb{N}_0$ \cite[sec. 5.8]{Fano-1961}.  For a linear, passive, time-invariant, narrow-fractional-bandwidth, two-port transmit-receive antenna system, the Friis transmission formula relates the average available receive-antenna power $\mathbb{S}$ to the average transmit-antenna accepted power $\mathbb{P}$  by  $\mathbb{S}\approx [G_T A_R {\cal L}/(4\pi r^2)]\mathbb{P}$, where $G_T$ and $A_R$ are the gain and effective area of the transmitting and polarization-matched receiving antenna, respectively, $r$ is the far-field separation distance between the antennas, ${\cal L}$ is any propagation loss factor (attenuation, scattering from objects) in addition to the $1/r^2$ loss.  The $-3$ dB bandwidth of the matched tuned transmitting antenna is used to approximate the finite rectangular bandwidth $\mathbb{B}$ in (\ref{sh}) \cite{S-H}.  It is assumed that the receiving antenna bandwidth adequately covers the $-3$ dB bandwidth of the transmitting antenna, and the factor $G_T A_R {\cal L}$ remains approximately  constant at its center-frequency value over the narrow $-3$ dB fractional bandwidth.  In all\\[-3mm]
\be{sh'}
\mathbb{C} \approx \mathbb{B} \log_2\{1 + [G_T A_R {\cal L}/(4\pi r^2)] \mathbb{P}/(\mathbb{N}_0\mathbb{B})\}.\\[-5mm]
\ee}
\par
The ingenious ``radio physicist'', H.A. Wheeler, was apparently the first to estimate the quality factor (the inverse of his ``radiation power factor'') of antennas by approximating the input impedance of both a series-inductor-tuned, top-and-bottom-loaded electric dipole and a series-capacitor-tuned, solenoidal magnetic dipole with lumped circuit elements \cite{Wheeler-1947}.  Although Wheeler's reasoning in \cite{Wheeler-1947} is not very transparent, his equations and results show that he uses the well-known capacitance $C=\eps_0 \kappa_e A/b$ and radiation resistance $R_e= Z_f (kb)^2/(6\pi)$ of a tuned electrically small straight-wire electric dipole of length $b$, top-and-bottom-loaded with two circular parallel plates each of area $A$.  He thus obtains the resonance series RLC quality factor
\be{Wh1}
 {Q^{\scriptscriptstyle \rm Wh}_e = \frac{1}{\w C R_e} = \frac{6\pi}{\kappa_e k^3 {\cal V}}}
\ee
where $k=2\pi/\lambda$ ($\lambda$ the free-space wavelength), $\eps_0$ and $Z_f$ are the free-space permittivity and impedance, $\kappa_e$ is a factor that accounts for the fringe electric fields that extend beyond the edges of the parallel-plate capacitor, and the volume ${\cal V}$ equals $A b$.
\par
Likewise, he uses the well-known inductance  $L \!\!\!\!=\!\!\!\! \mu_0 {\cal N}^2 A/(\kappa_m b)$ and radiation resistance $R_m \!\!=\!\! Z_f ({\cal N}kA)^2/(6\pi)$ of a tuned electrically small ${\cal N}$-turn solenoidal magnetic dipole of length $b$ and cross-sectional area $A$ to obtain the resonance series RLC quality factor 
\be{Wh2}
 {Q^{\scriptscriptstyle \rm Wh}_m = \w L/R_m = \frac{6\pi}{\kappa_m k^3 {\cal V}}} 
\ee
where $\mu_0$ is the free-space per\-me\-a\-bil\-ity, and $\kappa_m$ is a factor that accounts for the fringe magnetic fields that extend beyond the ends of the solenoid.
\par
For circular cylinders with height to diameter ratios of about unity, Wheeler estimates $\kappa_e \approx 3$ and $\kappa_m \approx 1.5$, so that $Q^{\scriptscriptstyle \rm Wh}_e$ and $Q^{\scriptscriptstyle \rm Wh}_m$ are not far from the minimum value of the quality factors found in \cite[sec. 6.2 with $\alpha_{^\{{\overset{e}{\scriptscriptstyle m}}\scriptscriptstyle^\}}\!\! \approx\! \kappa_{\scriptscriptstyle^\{\!{\overset{e}{m}}{\scriptscriptstyle^\}}}\scriptstyle \!\cal V$]{Y&G&J} for electric-current-produced electric and magnetic dipole antennas confined to these size circular cylinders.
\par
In his 1958 paper, Wheeler, using the same reactance/(radiation-resistance) technique (explained above) as in his 1947 paper, showed that electrically small spherical electric- and magnetic-dipole antennas having lossless magnetodielectric cores with frequency-independent relative permeability $\mu_{\rm rel}$ and permittivity $\eps_{\rm rel}$ excited by global electric currents had minimum quality factors $Q_m$ and $Q_e$, respectively, given by\\[-1.5mm]
\begin{subequations}
\lbl{Whme}
 \be{Whmea}
 {Q_m =\left(1+\frac{2}{\mu_{\rm rel}}\right)\frac{1}{(ka)^3}}\\[-1mm]
\ee
\be{Whmeb}
 {Q_e =\left(1+\frac{\eps_{\rm rel}}{2}\right)\frac{1}{(ka)^3}}\\[-1.5mm]
\ee
\end{subequations}
where $ka$ is the electrical size of the spherical antenna with radius $a$ \cite{Wheeler}, \cite{Kim-minQ}.
Wheeler observed that filling the antenna volume with a material of infinite magnetic permeability [$\mu_{\rm rel}\to\infty$ in (\ref{Whmea})] would reduce the stored magnetic energy inside the sphere to zero, enabling the Chu/Collin-Rothschild (CR) lower bound [see (\ref{QChu}) below] to be achieved theoretically for $ka\ll 1$. Despite the practical issues of internal resonances and producing the required low-loss magnetic material, 
this result provides an approach for reducing the internal energy and lowering the $Q$ of small 
magnetic-dipole antennas using permeable materials \cite{K&B&Y}. Wheeler offered no analogous method for 
reducing the internal stored electric energy of an electrically small spherical electric-dipole antenna (except for noting the unphysical frequency-independent zero-permittivity condition), but observed that for $\mu_{\rm rel}\eps_{\rm rel}=4$, the electrically small magnetodielectric spherical dipole antenna with equal power in the electric and magnetic dipoles would be self resonant. [Note that a high-permeability material filling the antenna volume is not equivalent to filling the 
volume with a perfect magnetic conductor (PMC) because the electric field, unlike the magnetic 
field, does not approach zero inside a solid lossless high-permeability core \cite{H&K&B}.]
\par
Chu's lower bound calculation was modified by Thal \cite{Thal} to account for the stored 
energy inside the bounding sphere when the antenna consists of global electric currents confined 
to the spherical surface surrounding an air core. The Thal lower bound on $Q$ for an electrically small electric-dipole antenna is a 
factor of about $1.5$ higher than the Chu/CR lower bound in (\ref{QChu}).  This result concurs with (\ref{Whmeb}) for $\eps_{\rm rel}=1$ and with the known performance of several optimized electrically small spherical electric-dipole antennas \cite{Best, S&P, S&B, Stuart}, and was generally thought for several years after Thal's 2006 paper to represent the best achievable performance for electric-dipole antennas \cite{H&C} (although Thal never said this in his paper).  
\par
However, the derivation in \cite{Y&S} suggested that a suitably driven thin shell of lossless permeable material surrounding an electric-dipole antenna can produce magnetic polarization currents that reduce the stored electric energy inside the antenna to obtain a $Q$ below the Thal lower bound.  In \cite{S&Y},  this idea was used to illustrate electric-dipole antenna designs incorporating thin shells of high relative magnetic permeability to attain $Q$'s close to the Chu/CR lower bound in (\ref{QChu}).  Kim \cite{Kim-minQ}, \cite{Kim} used global electric surface currents over high-permeability spherical magnetic shells with either a perfect electrical conductor or air inside to design electric- and magnetic-dipole antennas with quality factors close to the Chu/CR lower bound.  For spherical magnetic cores with loss tangents so large that the cores behave as PMC's, the quality factors of both electric and magnetic dipoles also approach the Chu/CR lower bound \cite{H&K&B}, \cite{HKB-J2014}.
\section{VSWR Impedance Bandwidth for Antennas}
The bandwidth of an antenna can be defined with respect to different parameters such as impedance, gain, beamwidth, or polarization.  Here we shall concentrate on input-impedance bandwidth, which is generally the main factor limiting the usable bandwidth of an electrically small antenna.  The input impedance of an electrically small antenna varies rapidly with frequency so as to limit the frequency range over which the antenna can be acceptably tuned and matched to its feed waveguide \cite{Jani}.
\par
If the impedance of an antenna is known as a function of frequency, one can directly determine the VSWR fractional bandwidth of the antenna.  Specifically, the fractional input-impedance VSWR bandwidth of a one-port feed-waveguide matched (often but not necessarily to 50 ohms), linear, passive, time-invariant antenna tuned to an isolated resonant or antiresonant frequency $\w$ is given by \cite{Y&B-2005}\\[-2mm]
\be{B1}
 { FBW_\beta(\w) \approx \frac{4\sqrt{\beta}\, R(\w)}{\w|Z'(\w)|},\;\;\beta = \frac{\alpha}{1-\alpha}=\frac{(s-1)^2}{4s}}\\[-1mm]
\ee
where $Z(\w)\!=\!R(\w)\!+\!jX(\w)$ is the input impedance of the antenna with $e^{j\w t}$ ($\w>0$) time dependence, $\alpha$ is the fractional power reflected at the edges of the band ($|\Gamma|^2=\alpha$, where $\Gamma(\w)$ is the antenna reflection coefficient)\footnote{The return loss is defined as $1/|\Gamma(\w)|^2$, so that $|\Gamma(\w)|^2$ in dB is the negative of the return loss in dB.}, and $s$ is the value of the VSWR at the edges of the band.  Primes denote differentiation with respect to $\w$.  This formula for bandwidth in terms of $\beta$ holds independently of the composition of the antenna and is accurate to the extent that $R(\w)/(\w |Z'(\w)|)$ does not change appreciably over the $\beta$-bandwidth.  Assuming that $Z'(\w) \neq 0$, the formula can always be made accurate if $\alpha$ (or equivalently $\beta$) is chosen small enough, that is, if the fractional reflected power is small enough.  For example, the formula (\ref{B1}) may not accurately predict the $-3$ dB bandwidth ($\alpha = 1/2$) yet give an accurate value for the $-10$ dB bandwidth ($\alpha = 1/10$).  Although $\alpha$ can range in value from zero to unity, it is typically desirable to have $\alpha \la 1/2$ and thus $\beta \la 1$ in order to prevent transmitter inefficiencies, distortion, and interference.
\par
If the feed waveguide characteristic resistance is not matched to the resistance of the antenna at the resonant/antiresonant frequency, it can be shown \cite{Pues} that for a given $\beta$ the fractional bandwidth in (\ref{B1}) can be increased by a maximum factor of $\sqrt{1+\beta}$, which is insignificant for $\beta\ll 1$.
\par
If $Z'(\w)=0$, there is no reflected power small enough to make (\ref{B1}) an accurate formula for the fractional bandwidth.  This exceptional case of $Z'(\w)=0$ [both $X'(\w)=0$ and $R'(\w)=0$] in addition to the tuned condition $X(\w)=0$ corresponds to the confluence of two or more resonances/antiresonances \cite{Stuart-2007}.  Although for $\beta$ small enough, the fractional bandwidth in (\ref{B1}) applies to regions of arbitrarily small nonzero spaced resonances and antiresonances, the $\beta$ required to ensure the accuracy of (\ref{B1}) for an isolated tuned resonance/antiresonance may be too small to define a useful bandwidth (depending on the application), and the more useful practical bandwidth may be the one that extends over a much larger untuned bandwidth where the magnitude squared of the reflection coefficient remains below a larger allowable value of $\alpha$.  Several ultrawideband antennas with reflection coefficients below $-10$ dB have been designed and built for antennas with electrical sizes as small as about unity  ($ka\approx 1$, where $a$ is the radius of the antenna's circumscribing sphere) at the lower end of their bands \cite{Stutzman,Wang,T&A}.  Also, the bandwidths of electrically larger ($ka \ga 1$) reflector antennas are limited only by the bandwidths of their feed antennas.
\subsection{Derivation of Antenna Fractional Bandwidth $FBW_\beta(\w)$}
The derivation of the fractional bandwidth expression in (\ref{B1}) begins with the definition of VSWR bandwidth for an antenna tuned at the frequency $\w$; namely as the difference between the two frequencies on either side of $\w$ at which the VSWR equals a constant $s$, or, equivalently, at which the magnitude squared of the reflection coefficient $|\Gamma(\w +\Delta\w_{\pm})|^2$ equals $\alpha = (s-1)^2/(s+1)^2$, provided the antenna is also matched at the frequency $\w$, that is, the characteristic impedance $Z_{\rm ch}$ of the single-mode feed waveguide equals the input resistance $R(\w)$ at the tuned frequency $\w$ [$Z_{\rm ch} = R(\w)$]. In other words, the matched tuned VSWR impedance bandwidth ($\Delta\w_+ -\Delta\w_-$), where ($\Delta\w_+ >0, \Delta\w_- < 0$), is determined by \cite{Y&B-2005}
\begin{align}
\lbl{FBW1}
 &\;|\Gamma(\w + \Delta\w_{\pm})|^2\!=\!\frac{X^2(\w\! +\! \Delta\w_{\pm})+[R(\w\!+\! \Delta\w_{\pm})-R(\w)]^2}{X^2(\w\! +\! \Delta\w_{\pm})+[R(\w\! +\! \Delta\w_{\pm})+R(\w)]^2}\nonumber\\  &\hspace{60mm}= \alpha. && 
\end{align}
\normalsize
Both $|\Gamma(\w)|^2$ and its derivative with respect to $\w$ are zero at $\w$.  Consequently, $|\Gamma(\w)|^2$ has a minimum at $\w$ for all $\w$ at which the antenna is tuned [$X(\w)=0$] and matched [$Z_{\rm ch}=R(\w)$].  This means that the matched VSWR bandwidth ($\Delta\w_+ -\Delta\w_-$), unlike the conductance or resistance bandwidth, is well-defined at all matched tuned frequencies (for small enough $\alpha$), that is, throughout both the antiresonant [$X'(\w)<0]$) and resonant [$X'(\w)>0]$ frequency ranges.%\footnote{Foster's reactance theorem tells us that $X'(\w)\ge 0$ if $X(\w)$ is produced by passive inductors and capacitors.  Therefore, the reactance $X(\w)$ in a series [$R(\w),X(\w)$] circuit must contain active elements \cite{Yaghjian-X}.  Although this violation of Foster's reactance theorem does not affect the derivations and results of the present paper, it can have consequences, especially for antenna energy considerations in the time domain.}
\par
Multiplying by the denominator of the quotient in (\ref{FBW1}), rearranging terms, expanding in a Taylor series about $\w$, and neglecting terms of $O(\Delta\w^3_\pm)$ yields \cite{Y&B-2005}
\be{FBW2}
 { |Z'(\w)|^2 \Delta\w^2_\pm \approx 4\beta R^2(\w),\quad \beta =\frac{\alpha}{1-\alpha}}
\ee
from which we find
\be{FBW3}
 { FBW_\beta(\w) = \frac{\Delta\w_+ -\Delta\w_- }{\w} \approx 2\Delta\w_+ \approx \frac{4\sqrt{\beta} R(\w)}{\w|Z'(\w)|}}
\ee
and thus the expression in (\ref{B1}).
\subsection{The Bandwidth Quality Factor $Q_Z(\w)$}
In view of the simplicity of the formula in (\ref{B1},\ref{FBW3}), an exact pragmatic impedance-based bandwidth quality factor $Q_Z(\w)$ can be defined as \cite{Y&B-2005}
 \be{B5}
 {Q_Z(\w) = \frac{\w|Z'(\w)|}{2R(\w)}}
\ee
which accurately relates to the inverse of the $\beta$-bandwidth as given by (\ref{B1},\ref{FBW3}) under the condition that $\beta$ is small enough that $R(\w)/(\w |Z'(\w)|)$ does not change appreciably over the $\beta$-bandwidth ($Z'(\w)\neq 0$).  The expression for $Q_Z(\w)$ in (\ref{B5}), which is evaluated at the center (resonant/antiresonant) frequency $\w$ of the band, does not depend on a coordinate origin for the fields of the antenna and it inherently ignores any extra lengths of transmission lines or ``surplus'' inductances and capacitances that may be present in the antenna \cite{Y&B-2005}, \cite[p. 176]{Karni}.  Since the center-frequency fractional bandwidth must always be less than $2$, the expressions in (\ref{B1},\ref{FBW3}) and (\ref{B5}) imply that $Q_Z(\w) > \sqrt{\beta_{\rm max}}\;$, where $\beta_{\rm max}$ is the largest value of $\beta$ for which (\ref{B1},\ref{FBW3}) is an accurate approximation [$Q_Z(\w) > 1$ for $\beta_{\rm max}=1$ (half-power bandwidth)].  However, as explained above, if the required $\beta_{\rm max}$ is very small, the $Q_Z(\w)$ bandwidth [$FBW_{\beta_{\rm max}}(\w) = 2\sqrt{\beta_{\rm max}}/Q_Z(\w)$] may be irrelevant to the practical effective bandwidth over a larger allowable value of $\beta$.  In general, (\ref{B1},\ref{FBW3}) and (\ref{B5}) combine to give
 \be{FB}
 {FBW_\beta(\w) \approx \frac{2\sqrt{\beta}}{Q_Z(\w)}}.
\ee
It follows from (\ref{FBW1}), (\ref{FBW2}), and (\ref{B5}) that for $\Delta\w$ within the $\beta$-bandwidth about the center frequency $\w$ for which (\ref{B1},\ref{FBW3}) is valid, the square of the magnitude of the reflection coefficient satisfies the equation\footnote{\lbl{Dcube}The right-hand side of (\ref{Gam}) ignores the $O[(\Delta\w/\w)^3]$ term because the odd terms do not change the fractional bandwidth; see Fig. \ref{fig5}.  In this sense, (\ref{Gam}) is valid up to $O[(\Delta\w/\w)^6]$.}
\be{Gam}
{|\Gamma(\w \pm \Delta\w)|^2 \approx \frac{[Q_Z(\w)\Delta\w/\w]^2}{1+[Q_Z(\w)\Delta\w/\w]^2}}
\ee
which approaches quadratic variation in $\Delta\w/\w$ as $\Delta\w/\w \to 0$.
\par
A useful formula for the matched tuned $Q_Z(\w)$ in terms of the antenna reactance $X_a(\w)$ \textit{before it is tuned with a lossless series capacitor or inductor} is given in \cite{Y&B-2005} as\footnote{Tuning the antenna with a lossless series (rather than parallel) reactance, which is assumed throughout, has the advantage of canceling the original reactance of the antenna without changing its original resistance.}
\bea{QZa}
Q_Z(\w) =\frac{\w}{2R(\w)}|Z'(\w)|\hspace{40mm}\nonumber\\
 = \frac{\w}{2R_a(\w)}\sqrt{[R_a'(\w)]^2 + [X_a'(\w)+|X_a(\w)|/\w]^2} 
\eea
where $R(\w)\!=\!R_a(\w)$ so that $R'(\w)\!=\!R'_a(\w)$.
For many very small ($ka \ll 1$) antennas, $X'_a(\w) \approx |X_a(\w)|/\w \gg |R_a'(\w)|$, so that
\be{QZa1}
 {Q_Z(\w) \approx \frac{|X_a(\w)|}{R_a(\w)}}
\ee
an engineering rule-of-thumb formula for quality factor that has its roots in K.S. Johnson's ``inductive purity'' (see Section \ref{Intro}).  The $Q_Z(\w)$ defined in (\ref{B5}) is exactly equal to the quality factor of a resonant series or antiresonant parallel RLC circuit with $R$ independent of frequency, that is, $Q_Z(\w) = \w L/R = 1/(\w CR)\;\;\mbox{or}\;\;Q_Z(\w)= R/(\w L)=\w CR$, respectively, with $R'(\w)=0$.   
\par
Even if $X'(\w) = 0$ over an extended frequency range, (\ref{FBW3}) and (\ref{FB}) give the fractional bandwidth for any one matched tuned frequency $\w$ within this range for $R'(\w) \neq 0$ and a small enough $\beta$.
\subsection{Bode-Fano Increase in Antenna Bandwidth\lbl{BFM}}
As explained above, provided $Q_Z(\w)\neq 0$, the formula in (\ref{FB}) for the fractional impedance bandwidth is always accurate at a small enough value of $\alpha$ (small enough value of $\beta$) for a single isolated resonance or antiresonance of a matched tuned antenna (no matter how small the nonzero separation between the ``isolated'' resonance/antiresonance and its neighboring resonance/antiresonance if $\beta$ is allowed to have indefinitely small values).  If an untuned antenna can be fed through a lossless circuit composed of capacitors and inductors that produce overlapping resonances/antiresonances, then Bode \cite[sec. 16.3]{Bode-1945} and Fano \cite[sec. I]{Fano} have proven that, with an unlimited number of resonances/antiresonances (that is, an unlimited number of capacitors and inductors), the fractional power reflected by the antenna can be made to equal a given constant ($|\Gamma|^2 = \alpha$) over the bandwidth and to rapidly approach (step-wise in the limit) unity ($100\%$ reflected power) outside the bandwidth to give a rectangular $|\Gamma(\w)|^2$ versus $\w$ reflected power curve.  This flat reflected power bandwidth, which is the largest possible bandwidth for any given value of $\alpha$, can be written as \cite[sec. 16.3]{Bode-1945},  \cite[sec. I]{Fano}
\be{BF2}
 { FBW^{\rm\scriptscriptstyle BF}_\beta(\w) =\frac{2\pi}{Q_Z(\w)\ln{(1/\alpha)}},\;\;\; \alpha = \frac{\beta}{1+\beta}}.
\ee
For half-power ($-3$ dB) fractional bandwidth, $\alpha = 1/2$, $\beta = 1$ and
 \be{BF3}
 {FBW^{\rm\scriptscriptstyle BF}_1(\w) = \frac{2\pi}{Q_Z(\w)\ln{(2)}} \approx \frac{9}{Q_Z(\w)}} 
\ee
a factor of about $4.5$ larger than the ordinary matched, series-capacitor-or-inductor tuned single-frequency half-power fractional bandwidth given by $FBW_1(\w) = 2/Q_Z(\w)$ in (\ref{FB}).  To avoid high losses and excessively complicated circuits, Bode-Fano tuning applied to electrically small antennas (ESAs, defined by $ka\la 1/2$) is often limited to two or three resonances/antiresonances that roughly halve this increase in half-power bandwidth \cite[fig.19]{Fano}.\footnote{The Bode-Fano limits on bandwidth given in (\ref{BF2}) apply  to RL and RC series and parallel circuits with $R$, $L$, and $C$ independent of frequency \cite{Fano}, \cite[sec. 5.9]{Pozar}.  Therefore, these limits can be applied to an untuned antenna at a frequency $\w$, provided the untuned antenna input impedance is well approximated by the impedance of an RL or RC series or parallel circuit with R, L, and C independent of frequency over a frequency band that extends to the chosen value of $\alpha$.  The largest allowable value of $\alpha$ is determined by the largest value of $\alpha$ for which $Q_Z(\w)$ in (\ref{FB}) accurately predicts the antenna fractional bandwidth.}
\par
\begin{figure}[h]
\vspace{-2mm}
\hspace{-4mm}
\includegraphics[height=1.6in,width=3.7in]{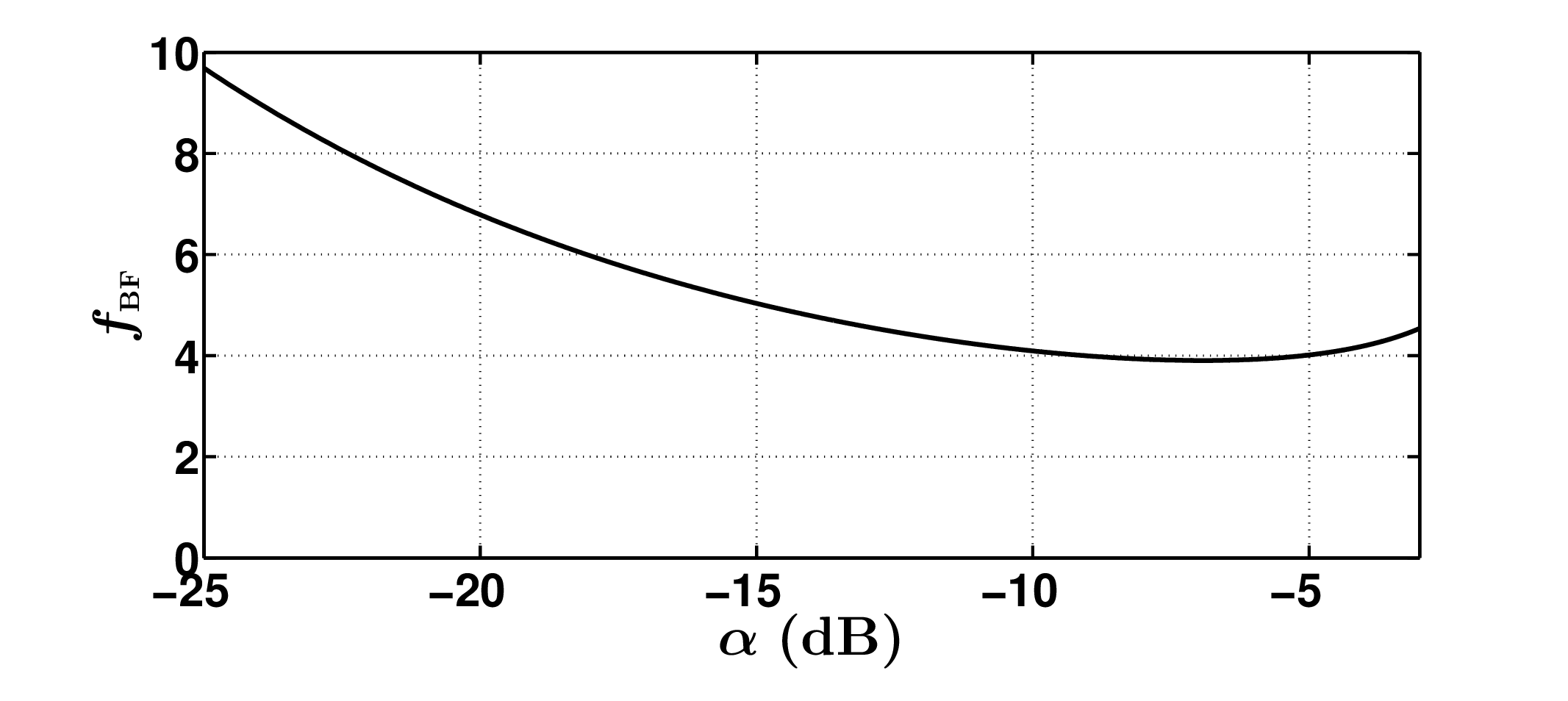}
\vspace{-7mm}
\caption{Maximum possible bandwidth factor increase $f_{\rm\scriptscriptstyle BF}$ (over ordinary single-resonance  bandwidth) produced by Bode-Fano multiple resonance/antiresonance tuning vs the bandwidth power-drop parameter $\alpha$.}
\label{BFfig}
%\mbox{}\\[-5mm]
\end{figure}

Dividing (\ref{BF2}) by (\ref{FB}) reveals that, in principle, Bode-Fano tuning can increase the ordinary single-resonance fractional bandwidth by a factor of
\be{BF4}
 { f_{\rm\scriptscriptstyle BF}(\alpha) = \frac{\pi\sqrt{(1-\alpha)/\alpha}}{\ln{(1/\alpha)}}\,}.
\ee
The same factor written in terms of the value $s$ of the VSWR bandwidth is given in \cite{Pues} and \cite[fig. 3.1b]{Jani}.  
This maximum theoretical improvement (\ref{BF4}) in fractional bandwidth produced by Bode-Fano tuning is plotted in Fig. \ref{BFfig} for the fractional reflected power-drop parameter $\alpha$ ranging from $-25$ dB  to $-3$ dB.  For the fractional reflected power-drop parameter $\alpha$ between about $-4$ dB and $-11$ dB,  Fig. \ref{BFfig} reveals that Bode-Fano multiple resonance/antiresonance tuning produces about a four-fold maximum possible increase in bandwidth over ordinary single resonance/antiresonance tuning.  With a realistic number of two or three resonances/antiresonances, the value of the improvement factor $f_{\rm\scriptscriptstyle BF}(\alpha)$ is roughly halved \cite[fig. 19]{Fano}, \cite[fig. 3.1b]{Jani}, leaving $f^{\rm\scriptscriptstyle realistic}_{\rm\scriptscriptstyle BF}(\alpha)\approx 2$ for $\alpha$ between about $-4$ dB and $-11$ dB.
\par
Although this realistic doubling of bandwidth is noteworthy for ESAs ($ka\la 1/2$), it should also be noted that the reflected power within the band never goes below about $\alpha$. (The average resistance of the antenna across the band is not perfectly matched to the characteristic resistance of the feed line, whereas the average reactance is usually negligible.)  The additional resonances and antiresonances increase the phase change (group delay) across the bandwidth;  thus Bode-Fano tuning can significantly increase the distortion of the transmitted signals of the antenna.  Also, Bode-Fano tuning increases the design and manufacturing complexity of the antenna as well as its ohmic loss and cost.  
\par
For wideband electrically larger ($ka \ga 1$) antennas, like horns and horn-fed reflectors, Bode-Fano tuning may be irrelevant because these antennas can inherently have low values of reactance $X$ oscillating about zero across their wide frequency bands \cite{Stutzman,Wang,T&A}.
\section{Conventional and Complex-Energy Quality Factors for RLC Circuit Models of Antennas}\lbl{accuracy}
A direct method, mentioned in Section \ref{IntroA}, for approximating the quality factor (and thus the bandwidth) of an antenna matched and tuned at a frequency $\w$ is to use the values of $X'(\w)$ and $R(\w)$ (assuming they are known at the tuned frequency $\w$) to obtain the values of $R$, $L$, and $C$ of the equivalent resonant ($X'(\w)>0$) or antiresonant ($X'<0$) RLC series or parallel circuit.  As an aside, although any impedance can be expressed mathematically as a reactance in series with a resistance, that is, $Z(\w)= R(\w) + jX(\w)$, if $X'(\w)<0$, then the Foster reactance theorem tells us that this series reactance $X(\w)$ cannot be produced by passive inductors and capacitors [even though the antenna and its total impedance $Z(\w)$ are passive, where passivity is defined in the time domain] \cite{Yagh-Passive}.
\subsection{Conventional RLC Quality Factors\lbl{CVQ}}
The conventional $Q_{\scriptscriptstyle\rm RLC}(\w)$ of the antenna modeled by an RLC circuit at the tuned frequency $\w$ is simply
\begin{subequations}
\lbl{QRLC}
 \be{QRLC1}
 {Q_{\scriptscriptstyle\rm RLC}(\w) = \frac{\w L}{R} = \frac{1}{\w CR}}
\ee
for a resonance, and
\be{QRLC2}
 {Q_{\scriptscriptstyle\rm RLC}(\w) = \frac{R}{\w L} = \w CR}
\ee
\end{subequations}
for an antiresonance, where $R$, $L$, and $C$ are approximated by different (constant) values at each tuned frequency $\w$.
\par
The conventional $Q_{\scriptscriptstyle\rm RLC}(\w)$ in (\ref{QRLC}) can be obtained from the average energy $W_{\scriptscriptstyle\rm LC}(\w)$ stored in the inductance $L$ and capacitance $C$ of the RLC circuit model, as well as the power dissipated $P_{\scriptscriptstyle\rm R}(\w)$ in $R$, by means of the IEEE definition \cite{IEEE} of quality factor\\[-3mm]
\begin{subequations}
\lbl{WRLC}
\be{WRLC1}
 {Q_{\scriptscriptstyle\rm RLC}(\w) = \frac{\w W_{\scriptscriptstyle\rm LC}(\w)}{P_{\scriptscriptstyle\rm R}(\w)}}\\[-2mm]
\ee
where\\[-3mm]
\be{WRLC2}
 \small{ W_{\scriptscriptstyle\rm LC}(\w)\! = \!\frac{1}{4}\bigg[\!\int\limits_C\!\eps_0 |\vE|^2 d{\cal V}\! +\!\int\limits_L \!\mu_0|\vH|^2 d{\cal V}\bigg]\! = \!\frac{1}{4}\!\!\left(L|I_{\scriptscriptstyle\rm L}|^2\! +\! C|V_{\scriptscriptstyle\rm C}|^2 \right)}
\ee
with $\vE$ and $\vH$ the electric and magnetic field in the capacitor and inductor, respectively,
and\\[-3mm]
\be{WRLC3}
{  P_{\scriptscriptstyle\rm R}(\w) = \frac{1}{2}R|I_{\scriptscriptstyle\rm R}|^2}\\[-2mm]
\ee
\end{subequations}
is the power accepted by the antenna.
With $I_{\scriptscriptstyle\rm L}$ the current in the inductor, $V_{\scriptscriptstyle\rm C}$ the voltage across the capacitor, and $I_{\scriptscriptstyle\rm R}$ the current in the resistor, (\ref{WRLC}) evaluates to give the conventional $Q_{\scriptscriptstyle\rm RLC}(\w)$ in (\ref{QRLC1}) and (\ref{QRLC2}) for series and parallel RLC circuits, respectively.
\par
A straightforward analysis of RLC circuits with $R$, $L$, and $C$ independent of frequency reveals that $Q_{\scriptscriptstyle\rm RLC}(\w)$ in (\ref{WRLC}) and the fractional VSWR bandwidth $FBW_\beta(\w)$ are related by
\be{QRBW}
 { FBW_\beta(\w)\! =\! \frac{\Delta\w_+ -\Delta\w_- }{\w}\! = \!\frac{2\sqrt{\beta}}{Q_{\scriptscriptstyle\rm RLC}(\w)}
\!= \!\frac{2\sqrt{\beta}}{Q_{Z\scriptscriptstyle\rm RLC}(\w)}}
\ee
with $Q_{\scriptscriptstyle\rm RLC}(\w)$ given in (\ref{QRLC}) and the bandwidth quality factor in (\ref{B5}) given by $Q_{Z\scriptscriptstyle\rm RLC}(\w) = \w |X'(\w)|/(2R)$ (with $R'(\w)$ assumed to be zero in the conventional RLC circuit model).  Notably, the formula in (\ref{QRBW}) is exact for all $\beta$ ($0<\alpha<1$) under the condition of $R$, $L$, and $C$ independent of frequency and without the requirement that the bandwidth is symmetric about the frequency $\w$, that is, $\Delta\w_-$ is not necessarily equal to $-\Delta\w_+$ in (\ref{QRBW}).   However, if $\sqrt{\beta}/Q_{\scriptscriptstyle\rm RLC} \ll 1$, then $\Delta\w_- \approx -\Delta\w_+$ and the bandwidth is approximately symmetric.
\par
For the general case in which $R'(\w) \neq 0$, the $Q_Z$ bandwidth quality factors defined in (\ref{B5}) for these resonant and antiresonant RLC circuit models of antennas are given by
\begin{subequations}
\lbl{QZ}
 \be{QZ1}
 {Q_Z(\w) = \frac{\w}{2R}\sqrt{R^{\prime 2}+4L^2}}
\ee
where $X^{\prime 2}=4L^2$
and\\[-2mm]
\be{QZ2}
 { Q_Z(\w) = \frac{\w}{2R}\sqrt{R^{\prime 2}+4R^4C^2}}
\ee
where $X^{\prime 2}=4R^4C^2$, respectively.  
\end{subequations}
A comparison of (\ref{QRLC}) with (\ref{QZ}) shows that
 \begin{subequations}
\lbl{QZRLC}
\be{QZRLC1}
 {Q_{\scriptscriptstyle\rm RLC}(\w) \approx Q_Z(\w)}\\[-3mm]
\ee
if and only if
\be{QZRLC2}
 {|R'(\w)| \ll |X'(\w)| \mbox{ say } \la |X'(\w)|/2}.
\ee
\end{subequations}
Since the inverse of $Q_Z(\w)$ in (\ref{FB}) gives an accurate fractional bandwidth of antennas for a small enough value of $\beta$ ($Q_Z(\w)\neq 0$), it follows from (\ref{QZRLC}) that the conventional RLC circuit-model quality factors $Q_{\scriptscriptstyle\rm RLC}(\w)$ in (\ref{QRLC}) predict a reasonably accurate VSWR fractional antenna bandwidth under the necessary and sufficient condition in (\ref{QZRLC2}) [assuming a small enough value of $\beta$ that $Q_Z(\w)\neq 0$ accurately predicts bandwidth].
\subsection{Complex-Energy RLC Quality Factors\lbl{CEQ}}
The conventional RLC circuit-model quality factors in (\ref{QRLC})--(\ref{WRLC}) do not include the effect of the antenna resistance changing with frequency, that is, the effect of $R'(\w)$.  This shortcoming can be remedied by using a generalized complex ``Q-energy'' defined in (\ref{CWfs}) of Section \ref{AFBQ} below (with $\vF\! =\! 0$) to account for the frequency-dependent resistor, and to obtain a generalized RLC circuit quality factor $Q^g_{\scriptscriptstyle\rm RLC}$, namely
\begin{subequations}
\lbl{QWRLC}
\be{QWRLC1}
  {Q^g_{\scriptscriptstyle\rm RLC}(\w) = \frac{\w |\cW^g_{\scriptscriptstyle\rm RLC}(\w)|}{P_{\scriptscriptstyle\rm R}(\w)}}
\ee
where the generalized complex Q-energy is given by
{ \bea{QWRLC2}
&&\hspace{-12mm}\cW^g_{\scriptscriptstyle\rm RLC}(\w)\! =\! \frac{1}{4}\bigg[\!\int\limits_C\!\eps_0|\vE|^2 d{\cal V}+\!\!\int\limits_L\! \mu_0|\vH|^2 d{\cal V}\nonumber\\&&\hspace{7mm}-j\!\int\limits_R\!\left(\sigma_e'(\w)|\vE|^2 +2\sigma_e(\w)\vE'\cdot\vE^*\right) d{\cal V} \bigg]\\
&&\hspace{-10mm}= \frac{1}{4}\left\{L|I_{\scriptscriptstyle\rm L}|^2 \!+\! C|V_{\scriptscriptstyle\rm C}|^2 \!-\!j\left[(1/R)'|V_{\scriptscriptstyle\rm R}|^2 +2V'_{\scriptscriptstyle\rm R}V^*_{\scriptscriptstyle\rm R}/R\right]\right\}\nonumber
\eea}\noindent
\end{subequations}
with $\sigma_e(\w)$  the frequency-dependent conductivity [so as to allow for a frequency-dependent $R(\w)$] of a cylindrical resistor with uniform axial electric field inside the cylinder.  The superscript $*$ denotes the complex conjugate.
Evaluation of the right-hand side of (\ref{QWRLC2}) for the series and parallel RLC circuits at the tuned frequency $\w$, then insertion into (\ref{QWRLC1}) gives\\[-2mm]
\be{QQRLC}
Q^g_{\scriptscriptstyle\rm RLC}(\w)\! =\! \frac{\w\sqrt{\stackrel{4L^2}{\scriptstyle 4R^4C^2}\stackrel{\;+\,R'^2(\w)}{}}}{2R(\w)}\! =\! \frac{\w|Z'(\w)|}{2R(\w)} \!=\! Q^g_{Z\scriptscriptstyle\rm RLC}(\w)
\ee
with the $4L^2$ and $4R^4C^2$ terms under the square root sign applying to the series and parallel RLC circuits, respectively.
This equation reveals the noteworthy result that the generalized complex Q-energy formulas give quality factors for RLC circuit models of resonant and antiresonant antennas that are exactly equal to the bandwidth quality factors $Q^g_{Z\scriptscriptstyle\rm RLC}(\w)$ in (\ref{B5}) obtained from the impedance derivatives $|Z'(\w)|$ for these models.
\section{Accurate Field-Based Quality Factors for Antennas}\lbl{AFBQ}
The bandwidth quality factor $Q_Z(\w)$ in (\ref{B5}) is robust in that it accurately determines the fractional bandwidth in (\ref{FB}) of the matched tuned antenna for a small enough power drop $\alpha$ (or, equivalently, $\beta$), it only requires a knowledge of $|Z'(\w)|$ and $R(\w)$ at the tuned center frequency $\w$ of the band, it is independent of the origin of the coordinate system of the antenna, and it is not affected by extra lengths of transmission lines or ``surplus'' inductances and capacitances innate or added to the antenna \cite{Y&B-2005}, \cite[p. 176]{Karni}.
Therefore, for any range of frequencies for which the input impedance $Z(\w)$ and resistance $R(\w)$ of an antenna are known analytically or computationally, one can rely solely on $Q_Z(\w) = \w|Z'(\w)|/[2R(\w)]$ ($Z'(\w)\neq 0$) as an accurate formula for antenna quality factor and its associated fractional bandwidth in (\ref{FB}).  And, indeed, $Q_Z(\w) = \w|Z'(\w)|/[2R(\w)]$ has become a widely used formula for predicting the quality factor and bandwidth of antennas.
\par
However, if the antenna current but not the voltage is known,\footnote{For example, many textbook linear and aperture antennas have the wire current or equivalent aperture surface current specified but not the voltage feeding the antenna \cite[chs. 4,8,12]{Balanis}.  For each antenna, an input current $I(\omega)$ must be defined.  Fortunately, however, both ${\cal W}$ and $P_{\rm R}$ in (31) contain a factor of $I^2(\omega)$ that cancels in the evaluation of ${\mathcal Q}$.} the input impedance as a function of frequency may not be readily available for an antenna to enable the evaluation of its quality factor and associated fractional bandwidth by means of $Q_Z(\w)$.  Moreover, it is desirable to determine expressions for the quality factor of an antenna directly in terms of its electric and magnetic fields.  Such field expressions for quality factor can indicate to the antenna designer ways in which the antenna can be modified to improve its bandwidth.  The field expressions for quality factor are especially amenable to theoretical analyses such as those for determining the lower bounds on the quality factor for differently shaped antennas \cite{Y&G&J}, and for maximum-gain antennas as in Section \ref{MG} below.  Field-based quality factors are also essential in computational optimization methods with electric-current expressions for the quality factors of antennas \cite{Gustafsson, Vandenbosch}.
\par
Since $Q_Z(\w)$ is the ``gold-standard'' antenna quality factor that is accurately related to fractional bandwidth (for a small enough $\beta$ and $Q_Z(\w)\neq 0$), it would be ideal to determine a general field-based quality factor $Q(\w)$ that equals $Q_Z(\w)$ (if it were available), especially since presently existing field-based quality factors (which are generally origin dependent\footnote{The previous field-based quality factors defined in \cite{Y&B-2005,Y_AWPL,Y&G&J} are origin independent only if the far-field pattern of the antenna satisfies the condition $\int_{4\pi}\vrh|\vF|^2 d\Omega = 0$, as does the far-field pattern of individual electric (TM) or magnetic (TE) spherical multipoles \cite{Y&B-2005}.} and subject to spurious energy contributions from extra lengths of transmission lines and surplus reactive elements) only approximate $Q_Z(\w)$ to an accuracy that is sometimes not acceptable. Also, it is often not feasible, in practice, to estimate the bandwidth-determining accuracy of the existing field-based quality factors applied to specific antenna designs.  Fortunately, formulas for robust field-based quality factors have been recently derived that equal $Q_Z(\w)$ \cite{Yaghjian-LACAP}.  The derivation of these robust field-based quality factors follows.
\begin{figure}[h]
\vspace{-4mm}
\hspace{1mm}
%\begin{center}
\includegraphics[height=1.4in,width=3.35in]{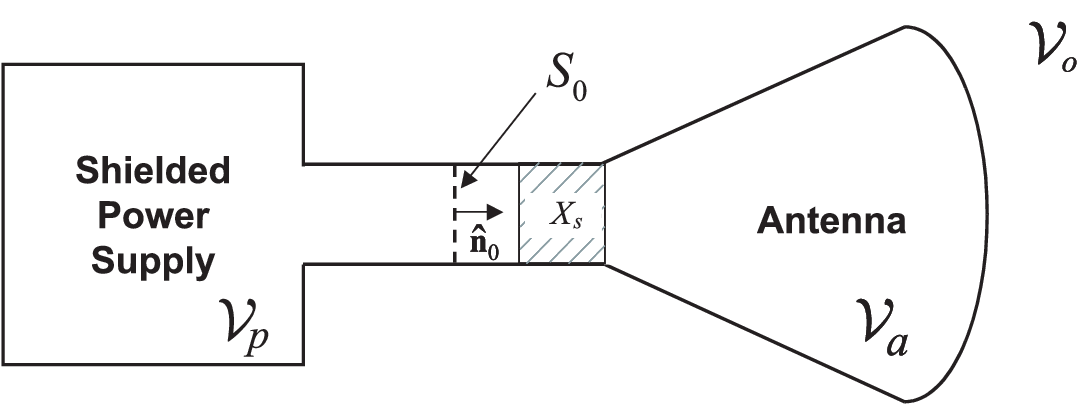}
%\end{center}
\vspace{-2mm}
\caption{Schematic of a general transmitting antenna, its matched single-mode feed waveguide, its shielded power supply ($\V_p$), and a series tuning reactance $X_s$.  The volume ${\cal V}_p$ is contained within the surface of the shielded power supply plus feed waveguide up to the reference plane $S_0$ (waveguide port).  The infinite volume outside the volume ${\cal V}_p$ is denoted by ${\cal V}_o$, which contains the material of the antenna ${\cal V}_a$.  The volume ${\cal V}_a$ of the antenna includes the element of the series tuning reactance $X_s$.}
\label{AntennaFig}
\end{figure}
%\vspace{-4mm}
%
\par
The derivation of the field-based quality factors relies on two fundamental theorems obtained for the electromagnetic fields obeying Maxwell's equations, namely the well-known complex Poynting theorem involving the Maxwellian fields, and a lesser known theorem similar to the Poynting theorem but involving the frequency derivatives of the fields as well as the fields themselves \cite{Y&B-2005}, \cite{Levis-1957}, \cite[sec. 8-5]{Harrington}, \cite{Fante-1969}.  These two theorems were applied in \cite{Y&B-2005} to the Maxwellian fields and their frequency derivatives for a general antenna shown schematically in Fig. \ref{AntennaFig} to obtain the following field expressions for the resistance $R(\w)$ \cite[eq.(64) with $P_A = |I|^2 R/2$]{Y&B-2005}, the reactance $X(\w)$ \cite[eq. (49)]{Y&B-2005}, and its frequency derivative $X'(\w)$ \cite[eq.(64)]{Y&B-2005}\\[-5mm]
\bea{Rf}
|I|^2 R(\w)\! = \!-\w\mbox{Im}\! \int\limits_{\V_o} \!\left(\vB \!\cdot\! \vH^* \!+ \!\vD\! \cdot\! \vE^*\right)d\V \!+\!\frac{1}{Z_f}\!\int\limits_{4\pi}\!|\vF|^2\, d\Omega\\[-4mm]\nonumber
\eea
[with the two integrals determining the dissipated and radiation antenna losses (resistances), respectively]\\[-3mm]
\be{Xf}
  |I|^2 X(\w)= \w\mbox{Re} \int\limits_{\V_o} \left(\vB \cdot \vH^* - \vD \cdot \vE^*\right)d\V\\[-4mm]
\ee
and\\[-5mm]
\bea{X'f}
&&\hspace{-3mm}|I|^2 X'(\w) = \lim_{r\to \infty}\bigg[\mbox{Re}\!\!\int\limits_{\V_o(r)}\!\!\left[(\w\vB)'\cdot\vH^* +(\w\vD)' \cdot\vE^*\right. \nonumber\\[-2mm] && \left.-\w\left(\vB^*\cdot\vH'+\vD^*\cdot\vE'\right)\right]d\V - 2\eps_0 r \int\limits_{4\pi}|\vF|^2\, d\Omega \bigg]\nonumber\\[-2mm]
&&\hspace{25mm}+\frac{2}{Z_f}{\rm Im}\int\limits_{4\pi}\vF'\cdot\vF^*\, d\Omega\\[-4mm]\nonumber
\eea
\textit{where the frequency derivatives of the fields are taken holding $I(\w)$ constant}.  [This does not mean that $I(\w)$ cannot change with frequency, only that at each frequency at which the derivatives of the fields are taken, the $I(\w)$ is treated as a constant at that frequency, that is, $d/d\w$ is a partial derivative $(\partial /\partial \w)_I$.]  Mathematical rigor requires that the volume $\V_o(r)$ be defined as the volume $\V_o$ capped by a sphere of radius $r\to\infty$.  The complex far electric field is defined by $\vF(\vrh)= \lim_{r\to\infty} r e^{jkr}\vE(\vr)$ and $\vE$, $\vD$, $\vB$, and $\vH$ are the usual Maxwellian field vectors with $Z_f= \sqrt{\mu_0/\eps_0}$ the impedance of free space.  The solid angle integration element $d\Omega$ is equal to $\sin{\theta}d\theta d\phi$ in the usual spherical coordinates.  The equalities in (\ref{Rf}), (\ref{Xf}), and (\ref{X'f}) are exact.  Note that taking the frequency derivative of (\ref{Xf}) does not result in an expression that is recognizably equal to the expression in (\ref{X'f}), even though the two are identically equal.
\par
Since the $Q_Z(\w)$ bandwidth quality factor requires $R'(\w)$ in addition to $X'(\w)$, take the frequency derivative of $R(\w)$ in (\ref{Rf}) to get\\[-6mm]
\bea{R'f}
|I|^2 R'(\w) = -\mbox{Im}\!\!\int\limits_{\V_o(r)}\!\!\big[(\w\vB)'\cdot\vH^* +(\w\vD)' \cdot\vE^*\hspace{5mm}\nonumber\\[-2mm]  \hspace{-5mm} -\w\left(\vB^*\cdot\vH'+\vD^*\cdot\vE'\right)\big]d\V 
+\frac{2}{Z_f}{\rm Re}\int\limits_{4\pi}\vF'\cdot\vF^*\, d\Omega.\\[-4mm]\nonumber
\eea
From (\ref{X'f}) and (\ref{R'f}), $Z'(\w)\!=\!R'(\w)\!+\! jX'(\w)$ can be found as\\[-5mm]
\bea{Z'f}
&&\hspace{-7mm}|I|^2 Z'(\w)\! =\! (-{\rm Im}\!+\!j{\rm Re}) \lim_{r\to \infty}\bigg[\!\!\int\limits_{\V_o(r)}\!\!\big[(\w\vB)'\!\cdot\!\vH^*\! +\!(\w\vD)'\! \cdot\!\vE^*\nonumber\\[-2mm]&&
-\w\left(\vB^*\cdot\vH'+\vD^*\cdot\vE'\right)\big]d\V - 2\eps_0 r \int\limits_{4\pi}|\vF|^2\, d\Omega \bigg]\hspace{5mm}\nonumber\\[-3mm]&&\hspace{17mm}
+\frac{2}{Z_f}\int\limits_{4\pi}\vF'\cdot\vF^*\, d\Omega.\hspace{5mm}\\[-4mm]\nonumber
\eea
With $Z'(\w)$ in (\ref{Z'f}) and $R(\w)$ in (\ref{Rf}) given in terms of the fields of the antenna, we can now define a robust field-based quality factor $Q(\w)$ as
\begin{subequations}
\lbl{BQF}
 \be{BQFf}
  {Q(\w) = \w|\cW(\w)|/P_{\rm\scriptscriptstyle R}(\w)}
\ee
where the complex Q-energy $\cW(\w)$ can be defined as
\bea{CWf}
&&\hspace{-6mm} \cW(\w) = \frac{1}{4} \lim_{r\to \infty}\bigg[\!\int\limits_{\V_o(r)}\!\left[(\w\vB)'\!\cdot\!\vH^*\! +\!(\w\vD)'\! \cdot\!\vE^*\right. \nonumber\\[-2mm] && \left.-\w\left(\vB^*\cdot\vH'+\vD^*\cdot\vE'\right)\right]d\V - 2\eps_0 r \int\limits_{4\pi}|\vF|^2\, d\Omega \bigg]\nonumber\\[-2mm]
&&\hspace{25mm}-\frac{j}{2Z_f}\int\limits_{4\pi}\vF'\cdot\vF^*\, d\Omega\\[-3mm]\nonumber
\eea
and
\bea{PRf}
&&\hspace{-8mm}P_{\rm\scriptscriptstyle R}(\w) = \frac{1}{2}|I|^2R(\w)\\&& = -\frac{\w}{2}\mbox{Im}\! \int\limits_{\V_o} \!\left(\vB \cdot \vH^* + \vD \cdot \vE^*\right)d\V +\frac{1}{2Z_f}\!\int\limits_{4\pi}\!|\vF|^2\, d\Omega.\nonumber\\[-5mm]\nonumber
\eea
\end{subequations}
\par
An important feature of the field-based quality factor in (\ref{BQF}) is that it always predicts an accurate fractional bandwidth given by\\[-5mm]
\be{ABW}
FBW_\beta(\w) \approx \frac{2\sqrt{\beta}}{Q(\w)}\\[-2mm]
\ee
($Q(\w)\neq 0$) for a small enough bandwidth power drop $\alpha = \beta/(1+\beta)$ because $Q(\w)$ would exactly equal the input-impedance quality factor $Q_Z(\w)$ if the input impedance of the linear, passive, time-invariant antenna were available (that is, $V$ as well as $I$ defined for the antenna).  This field-based $Q(\w)$, like $Q_Z(\w)$, is origin independent and unaffected by any extra lengths of transmission lines or ``surplus'' inductances and capacitances\footnote{The stored energy in an extra length of transmission line attached to a resistor will be canceled by the $\sigma_e \vE'\cdot\vE$ integral in the Q-energy $\cW(\w)$ [see, for example, (\ref{CWfs})] because the current through the resistor is not in phase with the input current, which is held constant  in taking the frequency derivative of the electric field, and thus $\vE'$ in the resistor is not zero, but ensures that the $\sigma_e\vE'\cdot\vE$ integral cancels the stored energy in the extra line.  A similar argument applies to any surplus inductances and capacitances.} that may be present in the antenna \cite{Y&B-2005}, \cite[p. 176]{Karni}.  It seems quite remarkable (notwithstanding the complex Poynting theorem and its related theorem used to derive it) that the frequency derivative of the input impedance time $|I|^2$ on the left-hand side of (\ref{Z'f}) (if it is available) equals exactly the integrals of the fields of the antenna on the right-hand side of (\ref{Z'f}).  From (\ref{Z'f}) and (\ref{CWf}), one finds the simple relationship between $Z'(\w)$ and the complex Q-energy $\cW(\w)$\\[-3mm]
\be{ZpW}
\cW(\w) = -j\frac{Z'(\w)|I|^2}{4}\\[-2mm]
\ee
if $Z'(\w)$ is available.  One can define complex quality factors as
\be{ZpWc}
 {\cQ(\w) = \cQ_Z(\w) = \frac{\w \cW(\w)}{P_{\rm\scriptscriptstyle R}} = -j\frac{\w Z'(\w)}{2R(\w)}}\\[-2mm]
\ee
since $P_{\rm\scriptscriptstyle R}=|I|^2R/2$.\\[-4mm]
\par
The most general frequency dispersive but spatially nondispersive linear constitutive relations are the bianisotropic ones given as\\[-5mm]
\begin{subequations}
\label{Bian}
 \be{Biana}
 {\vD =\dep(\w) \cdot \vE + \dtau(\w) \cdot \vH}
\ee\noindent
\be{Bianb}
 {\vB =\dmu(\w) \cdot \vH + \dnu(\w) \cdot \vE}\\[-2mm]
\ee\noindent
\end{subequations}\noindent
with $\dep(\w)$, $\dmu(\w)$, and [$\dnu(\w),\dtau(\w)$] the permittivity dyadic, the permeability dyadic, and the magnetoelectric dyadics, respectively.  These bianisotropic constitutive relations can be substituted into (\ref{BQF}) to obtain the most general expressions for the field-based quality factor for linear frequency dispersive but spatially nondispersive antenna material, namely\\[-5mm]
\begin{subequations}
\lbl{biBQFs}
  {\bea{biCWfs}
&&\hspace{-6mm} \cW(\w) = \frac{1}{4} \lim_{r\to \infty}\bigg[\!\int\limits_{\V_o(r)}\big[\vH^*\cdot(\w\dmu)'\cdot\vH +\vE^*\cdot(\w\dep)'\cdot\vE \nonumber\\[-5mm] && \hspace{30mm}  +\vH^*\cdot(\w\dnu)'\cdot\vE +\vE^*\cdot(\w\dtau)'\cdot\vH \nonumber\\ && \hspace{5mm}  +\w\vH^*\cdot(\dmu-\dmu^*_\mathfrak{t})\cdot\vH' + \w\vE^*\cdot(\dep-\dep^*_\mathfrak{t})\cdot\vE'\nonumber\\  && \hspace{5mm} +\w\vH^*\cdot(\dnu-\dtau^*_\mathfrak{t})\cdot\vE' + \w\vE^*\cdot(\dtau-\dnu^*_\mathfrak{t})\cdot\vH' \big]d\V 
 \nonumber\\ &&\hspace{6mm}- 2\eps_0 r \int\limits_{4\pi}|\vF|^2\, d\Omega \bigg]-\frac{j}{2Z_f}\int\limits_{4\pi}\vF'\cdot\vF^*\, d\Omega\\[-8mm]\nonumber
\eea}
\bea{biPRfs}
 &&\hspace{-8mm}P_{\scriptscriptstyle R}(\w) = \frac{1}{2}|I|^2R(\w)\nonumber\\&& \hspace{-5mm}= -\frac{\w}{4} \int\limits_{\V_o}\big[\vH^*\cdot(\dmu-\dmu^*_\mathfrak{t})\cdot\vH + \vE^*\cdot(\dep-\dep^*_\mathfrak{t})\cdot\vE\nonumber\\&&  \hspace{10mm}+ \vH^*\cdot(\dnu-\dtau^*_\mathfrak{t})\cdot\vE + \vE^*\cdot(\dtau-\dnu^*_\mathfrak{t})\cdot\vH \big] d\V \nonumber\\ && \hspace{30mm} +\frac{1}{2Z_f}\int\limits_{4\pi}|\vF|^2\, d\Omega\\[-5mm]\nonumber
\eea
\end{subequations}
where the subscript $\mathfrak{t}$ indicates the transpose dyadic.  All the terms containing a minus sign between two dyadics are zero except in lossy antenna material.  The difference between the complex Q-energy in (\ref{biCWfs}) and the complex Q-energy defined previously, for example, in \cite[eq. (10)]{Y&G&J}, is the complex far-field term $-[j/(2Z_f)]\int_{4\pi}\vF'\cdot\vF^* d\Omega$ in (\ref{biCWfs}), as well as the terms with a minus sign between two dyadics, these latter terms being zero except in lossy antenna material.
\par
For scalar frequency dispersive lossy permittivity and permeability (that includes electric and magnetic conductivity as well as their real parts, denoted by the subscript $r$)\\[-3mm]
\be{scpp}
\eps(\w)\! =\! \eps_r(\w) \!-\! j\sigma_e(\w)/\w\,,\;\;\; \mu(\w)\! = \!\mu_r(\w) \!-\! j\sigma_m(\w)/\w\\[-1mm]
\ee
and (\ref{CWf}, \ref{biCWfs}) reduce to\\[-5mm]
\begin{subequations}
\lbl{BQFs}
\bea{CWfs}
&&\hspace{-6mm} \cW(\w) = \frac{1}{4} \lim_{r\to \infty}\bigg[\!\int\limits_{\V_o(r)}\left[(\w\mu)'|\vH|^2 +(\w\eps)'|\vE|^2\right. \nonumber\\[-3mm] && \left.-2j\left(\sigma_m\vH'\cdot\vH^* + \sigma_e\vE'\cdot\vE^*\right)\right]d\V - 2\eps_0 r \int\limits_{4\pi}|\vF|^2\, d\Omega \bigg]\nonumber\\[-3mm]
&&\hspace{25mm}-\frac{j}{2Z_f}\int\limits_{4\pi}\vF'\cdot\vF^*\, d\Omega\\[-7mm]\nonumber
\eea
\bea{PRfs}
&&\hspace{-8mm}P_{\rm\scriptscriptstyle R}(\w) = \frac{1}{2}|I|^2R(\w)\\&& \hspace{-5mm}= \frac{1}{2}\! \int\limits_{\V_o} \!\left(\sigma_m|\vH|^2 + \sigma_e|\vE|^2\right)d\V +\frac{1}{2Z_f}\!\int\limits_{4\pi}\!|\vF|^2\, d\Omega.\nonumber\\[-5mm]\nonumber
\eea
These equations (\ref{BQFs}) were used in Section \ref{CEQ} to obtain the exact generalized quality factors $Q^g_{\scriptscriptstyle\rm RLC}(\w)$  for resonant and antiresonant series and parallel RLC circuits, respectively.
\end{subequations}
\par
From (\ref{ZpW}), the real and imaginary parts of the expressions for the complex Q-energies in (\ref{biCWfs}) and (\ref{CWfs}) correspond to the contributions to $Z'(\w)$ from $X'(\w)$ and $R'(\w)$, respectively.  The  expressions for the stored energies $W$ used in \cite{Y&B-2005} equal the real parts of $\cW$ in (\ref{biCWfs}) and (\ref{CWfs}) without the terms with the primed fields.
\section{Revised Q Lower Bounds for Spherical Modes\lbl{RLB}}
In his classic 1948 paper, Chu found ladder networks of frequency independent resistors ($R$), inductors ($L$), and capacitors ($C$) that produce the input impedances derived from an integration of the complex Poynting theorem [see (\ref{Rf})-(\ref{Xf})] applied to the fields of the Stratton \cite[ch. 7]{Stratton} TM and TE spherical modes at a free-space radius $r=a$ \cite[eqs. (5)--(7)]{Chu}.\footnote{The eigenfunctions of the time-harmonic vector wave equation in spherical coordinates are more often referred to as ``spherical wave functions'' or just ``spherical waves'' in the physics literature rather than the more common engineering term ``spherical modes'' used herein.  The term ``spherical multipole'' is usually used to emphasize the unique functional dependence of the individual spherical modes.} Then assuming that the energy contributions to the quality factors from inside the sphere of radius $a$ are zero (except for energy in the lossless tuning elements), he derives analytically the average energy stored in the inductors and capacitors, and the average energy dissipated in the resistors, then determines the minimum circuit quality factors for the tuned spherical modes from twice the electric (magnetic) stored energy for the TM (TE) modes.  To simplify the evaluation of the minimum quality factors, Chu made the approximation of representing the input impedance of each spherical mode by an RLC equivalent circuit (with R, L, and C independent of frequency).  Nonetheless, the ladder networks in Chu's paper \cite{Chu} allow for the exact evaluation of his ladder-network tuned-circuit stored energy and power radiated, and thus the ladder-network stored-energy quality factors of the spherical modes.  It turns out that the TM and TE spherical modes have the same Chu quality factors even though they have different impedances.
\par
Sixteen years later, Collin and Rothschild, in their classic 1964 paper \cite{C&R}, evaluated the minimum quality factors for the Stratton \cite[ch. 7]{Stratton} TM and TE spherical modes by directly integrating the reactive energy in the free space outside a radius $a$ of the modes, with the reactive energy defined as the total energy in the fields less the radiated energy.  (Notwithstanding statements to the contrary \cite{Levis-1957}, the $1/r$ decaying radiated fields propagate at the speed of light in the free space outside the sources, since a Wilcox expansion of the fields in powers of $1/r$  readily shows that all the terms in the expansion propagate at the speed of light in the free space outside the sources \cite[p. 405]{Stratton}, \cite{Wilcox} .)  Quite remarkably, this direct integration of judiciously defined reactive energy led to the same ladder-network minimum quality factors determined earlier by Chu.  Actually, Collin and Rothschild proved that their direct-integration quality factors were equal to Chu's ladder-network quality factors only for the first three degree spherical modes ($n=1,2,3$).  Recently, Murray and Iyer \cite{M&I} have used mathematical induction to prove the equality of the Chu and Collin-Rothschild quality factors for every value of the degree $n$ of a spherical mode.  For the $n=1$ electric (TM) and magnetic (TE) dipole spherical modes, the Chu/CR minimum quality factors are given by the same simple well-known expression\footnote{Actually, Chu's paper does not explicitly contain the dipole lower-bound expression in (\ref{QChu}), although Chu says that it would not be difficult to calculate.}
\be{QChu}
Q_{1\rm Chu}(ka) = Q_{1\rm CR}(ka) = \frac{1}{(ka)^3} + \frac{1}{ka}.
\ee
\par
Notwithstanding the usefulness of this formula, especially for $ka \la 1$, it is not as accurate as the $Q_{nZ}(ka)$ quality factors for the $n=1$ spherical modes because the Chu field-based quality factors omit the $\vE'$ integrations in (\ref{CWfs}) in the resistance of the Chu ladder networks, and the CR integrations omit the $\vF'$ terms in (\ref{CWfs}), the terms needed to bring the field-based quality factors equal to $Q_{nZ}(ka)$, that is, equal to robust quality factors $Q_n(ka)$ that accurately determine the fractional bandwidth from the formula in (\ref{ABW}).  The CR formulation also has ($r-a$) replacing $r$ in (\ref{CWfs}).  Therefore, we shall evaluate $Q_{nZ}(ka)$ for spherical modes in terms of their fields to obtain more accurate field-based minimum quality factors $Q_n(ka)$ for any value of degree $n$ and, in particular, for $n=1$.  The $Q_{nZ}(ka)$ quality factors for spherical modes were also evaluated by Gustafsson and Nordebo in \cite{G&N-2006}.\footnote{\lbl{CRfoot}Specifically, Collin and Rothschild performed the integrations for $\cW$ in (\ref{CWfs}), the accurately derived complex Q-energy, without the primed-field integrations and using a range for their volume integrations from $r=a$ to $r\to\infty$ (with $\eps = \eps_0$ and $\mu = \mu_0$) so that a factor of $(r-a)$ results in their expression corresponding to (\ref{CWfs}) instead of $r$.  The antenna sources in ${\cal V}_a$ of Fig. \ref{AntennaFig} for the minimum-Q spherical modes of Chu and Collin-Rothschild can be assumed to be equivalent electric and magnetic (magnetization) surface currents in an infinitesimally thin shell at $r=a$ fed from inside the sphere by a waveguide voltage $V$ and current $I$.  These surface currents generate the exterior fields ($r>a^+$) of the spherical modes while giving zero fields inside ($r<a^-$).  The stored energy in the shell of surface currents is assumed to approach zero as the thickness of the shell approaches zero.  For electrically small antennas with a required phase difference between the currents producing the spherical multipoles, such as maximum-directivity TM-plus-TE modes (see Section \ref{MG}), Thal \cite{Thal-2009} has shown that tuning elements are required that will increase the stored energy and thus increase the quality factor ($Q = Q_Z$); for example, approximately doubling the stored energy and quality factor for electrically small electric-plus-magnetic Huygens dipoles with a directivity of $3$ ($4.77$ dB).  However, dispersive tuning could be used to obtain the required phase difference without adding to the Q-energy or changing the directivity if the accompanying loss (which lowers the gain) and smaller bandwidth power drop can be tolerated (see Section \ref{DT}).}    
\subsection{Impedance-Based $Q_{nZ}(ka)$ for Spherical Modes}
From Chu's paper \cite{Chu}, the untuned input impedance $Z_n^{\scriptscriptstyle\rm TM}$ of an $n$th degree TM spherical mode for any value of order $m$ ($-n\le m \le n$) is given by\footnote{Historically, the $n$ in the spherical Hankel function is referred to as its order, whereas the corresponding $n$ in a spherical mode is referred to as its degree \cite{Jesper}.}
\begin{subequations}
\lbl{ZRXTM}
\be{ZTM}
 { Z_n^{\scriptscriptstyle\rm TM}(ka) = jZ_f\frac{[ka h^{(2)}_n(ka)]'}{ka h^{(2)}_n(ka)}}
\ee
with the real part $R_n^{\scriptscriptstyle\rm TM}(ka)$  and the imaginary part $X_n^{\scriptscriptstyle\rm TM}(ka)$ given by
\be{RTM}
 {R_n^{\scriptscriptstyle\rm TM}(ka) = \frac{Z_f}{[ka |h^{(2)}_n(ka)|]^2}}
\ee
and
\be{XTM}
X_n^{\scriptscriptstyle\rm TM}(ka) = Z_f\Bigg[\frac{1}{ka} +\frac{\big[|h^{(2)}_n(ka)|^2\big]'}{2|h^{(2)}_n(ka)|^2}\Bigg]\\[-2mm]
\ee
\end{subequations}
where $h^{(2)}_n(ka)$ is the spherical Hankel function of the second kind and the primes are now the derivatives with respect to $ka$.   Incidentally, there is no total reactance in the radiation fields because their magnetic energy is equal to their electric energy.
The matched tuned $Q_{nZ}^{\scriptscriptstyle\rm TM}(ka)$ can be determined from (\ref{QZa}) and (\ref{ZRXTM}) as
 \be{QZnTM}
 {Q^{\scriptscriptstyle\rm TM}_{nZ}(ka)\! =\! \frac{ka}{2R_n^{\scriptscriptstyle\rm TM}}\sqrt{[R_n^{{\scriptscriptstyle\rm TM}\,\prime}]^2 \!+\! [X_n^{{\scriptscriptstyle\rm TM}\,\prime}\!+\!|X_n^{\scriptscriptstyle\rm TM}|/(ka)]^2}}.
\ee
Because the evaluation of the right-hand side of (\ref{QZnTM}) is straightforward with the help of (\ref{ZRXTM}) but tedious for arbitrary $n$, we shall show the results for only the $n\!=\!1$ TM spherical modes (electric dipoles), namely
{ \bea{QZ1TM}
&&\hspace{-10mm}Q^{\scriptscriptstyle\rm TM}_{1Z}(ka) = \frac{ka}{2R_1^{\scriptscriptstyle\rm TM}}\sqrt{[R_1^{{\scriptscriptstyle\rm TM}\,\prime}]^2 + [X_1^{{\scriptscriptstyle\rm TM}\,\prime}+|X_1^{\scriptscriptstyle\rm TM}|/(ka)]^2}\nonumber\\
&& \hspace{5mm}= \frac{1}{(ka)^3}\frac{\sqrt{1+4(ka)^2 + 4(ka)^4 + (ka)^6}}{[1+(ka)^2]}\nonumber\\
&& \hspace{3mm}\stackrel{ka\ll 1}{=}\frac{1}{(ka)^3}+\frac{1}{ka}-ka+O[(ka)^3].
\eea}\noindent
A comparison of (\ref{QZ1TM}) with (\ref{QChu}) shows that the exact lower bound (minimum $Q_Z = Q$) for the electric dipole is smaller in value than the traditional Chu/CR lower bound, although for $ka \la 0.5$ the difference is inconsequential.  Also, the exact lower bound in (\ref{QZ1TM}) goes as $1/(ka)^2$ for $ka\gg 1$ rather than as $1/(ka)$ for the Chu/CR lower bound, a change that can strongly affect the value of the quality factors of electrically large ($ka\gg 1$) antennas as, for example, in (\ref{HGQA}) below compared to (\ref{HGQ}) and (\ref{HGQF}).
\par
A similar exercise with the $n$th degree TE spherical modes can begin with the reciprocal relationship between the untuned input impedances of the Chu TE and TM spherical-mode input impedances, that is
\be{ZTE}
 { Z_n^{\scriptscriptstyle\rm TE}(ka) = \frac{Z_f^2}{Z_n^{\scriptscriptstyle\rm TM}(ka)} = -jZ_f\frac{ka h^{(2)}_n(ka)}{[ka h^{(2)}_n(ka)]'}\,}.
\ee
Because of this reciprocal relationship, the untuned quality factors for the TE spherical modes can be shown to be identical to the untuned quality factors for the TM spherical modes, that is\\[-3mm]
{ \bea{QZunt}
\widetilde{Q}_{nZ}(ka)=\widetilde{Q}^{\scriptscriptstyle\rm TE}_{nZ}(ka)=\widetilde{Q}^{\scriptscriptstyle\rm TM}_{nZ}(ka)\hspace{11mm}\nonumber\\  = \frac{ka|Z_n^{{\scriptscriptstyle\rm TM} \prime}(ka)|}{2R_n^{\scriptscriptstyle\rm TM}(ka)} = \frac{ka|Z_n^{{\scriptscriptstyle\rm TE} \prime}(ka)|}{2R_n^{\scriptscriptstyle\rm TE}(ka)}
\eea}\noindent
where the overtildes denote the untuned quality factors.  %{\color{black}[In (\ref{TMpTE'}), the self-tuned $Q_Z$ quality factors of the TM-plus-TE spherical modes are given as
%{  \be{TMpTE}
%Q^{\rm\scriptscriptstyle TMTE}_{nZ}(ka) = |1+\gamma_n| \widetilde{Q}_{nZ}(ka)/2
%\ee
%with the complex tuning factor $\gamma_n$ given in (\ref{gTMTE}) below.  
%The $Q^{\scriptscriptstyle\rm TMTE}_{nZ}(ka)$ are nearly equal to $Q_{n\rm F}(ka)/2 \approx Q_{n\rm Chu}(ka)/2$ for $ka\la 1$, where $|1+\gamma_n|\approx 2$, and the Fante $Q_{n\rm F}(ka)$ is given in (\ref{Fqf}) below.]}
\par
However, the tuned TE quality factors given by the same equation as (\ref{QZnTM}) but with the superscripts TM replaced by TE, that is\\[-3mm]
 \be{QZnTE}
 {Q^{\scriptscriptstyle\rm TE}_{nZ}(ka) = \frac{ka}{2R_n^{\scriptscriptstyle\rm TE}}\sqrt{[R_n^{{\scriptscriptstyle\rm TE}\,\prime}]^2 + [X_n^{{\scriptscriptstyle\rm TE}\,\prime}+|X_n^{\scriptscriptstyle\rm TE}|/(ka)]^2}}
\ee
are not equal to the tuned $Q^{\scriptscriptstyle\rm TM}_{nZ}(ka)$ in (\ref{QZnTM}) because $Z_n^{\scriptscriptstyle\rm TE}(ka) \neq Z_n^{\scriptscriptstyle\rm TM}(ka)$ and thus the series tuning reactances and radiation resistances have different values.  This difference in the tuned TM and TE $\,Q_{nZ}$ lower-bound quality factors stands in contrast to the tuned TM and TE Chu/CR lower bounds, which are equal, that is
\be{Chueq}
 { Q_{n\rm Chu}^{\scriptscriptstyle\rm TE}(ka)=Q_{n\rm Chu}^{\scriptscriptstyle\rm TM}(ka) =Q_{n\rm Chu}(ka)=Q_{n\rm CR}(ka)}
\ee
with $Q_{1\rm Chu}^{\scriptscriptstyle\rm TE}(ka)\!=\!Q_{1\rm Chu}^{\scriptscriptstyle\rm TM}(ka)\!=\! Q_{1\rm Chu}(ka)\!=\! Q_{1\rm CR}(ka)$ for $n=1$ given in (\ref{QChu}).
\par
Evaluation of $Q^{\scriptscriptstyle\rm TE}_{1Z}(ka)$ from (\ref{QZnTE}) yields
{ \bea{QZ1TE}
Q^{\scriptscriptstyle\rm TE}_{1Z}(ka)\!\!\!\!\!\!\!\!&=&\!\!\!\!\!\!\!\!\frac{1}{(ka)^3}\frac{\sqrt{1\!-\!2(ka)^4 \!+\! 4(ka)^6 \!-\! 3(ka)^8\! +\!(ka)^{10}}}{[1-(ka)^2 + (ka)^4]}\nonumber\\ \hspace{5mm}\!\!\!\!\!\!\!&\stackrel{ka\ll 1}{=}&\!\!\!\!\!\frac{1}{(ka)^3}+\frac{1}{ka}-ka+O[(ka)^3].
\eea}\noindent
A comparison of (\ref{QZ1TE}) with (\ref{QChu}) shows that the exact lower bound (minimum $Q_Z = Q$) for the magnetic dipole is smaller in value than the traditional Chu/CR lower bound, although for $ka \la 0.5$ the difference is inconsequential.  Also, the exact lower bound in (\ref{QZ1TE}) goes as $1/(ka)^2$ for $ka\gg 1$ rather than as $1/(ka)$ for the Chu/CR lower bound, {\color{black}a change that can strongly affect the value of the quality factors of electrically large ($ka\gg 1$) antennas as, for example, in (\ref{HGQA}) compared to (\ref{HGQ}) and (\ref{HGQF}).}
\par
Interestingly, the approximate $n=1$ quality factor used by Chu \cite{Chu}, which Fante denotes as $Q_1^{\scriptscriptstyle\rm CHU}(ka)$, is related to $Q_{1Z}^{\scriptscriptstyle\rm TM,TE}(ka)$ in (\ref{QZ1TM}) and (\ref{QZ1TE}) by \cite{Fante-1969}
\be{FanteEq}
  {Q_1^{\scriptscriptstyle\rm CHU}(ka) = Q_{1Z}^{\scriptscriptstyle\rm TM,TE}(ka) + O[(ka)^3].}
\ee
In other words, the approximate $Q_1^{\scriptscriptstyle\rm CHU}(ka)$ lower bound used by Chu in his paper agrees more closely with the ``exact'' $Q_{1Z}^{\scriptscriptstyle\rm TM,TE}(ka)$ lower bounds than does the actual $Q_{\rm 1Chu}(ka)=Q_{\rm 1CR}(ka)$ ladder-network lower bound (genius lucks out).
\par
The input impedances $Z_n^{\scriptscriptstyle\rm TM}(ka)$ and $Z_n^{\scriptscriptstyle\rm TE}(ka)$ in series will be self-tuned at a frequency $\w_0$, that is, $k_0a$, if one of the modes, say the TE mode, is excited by a constant multiplying factor $\alpha_n$ such that the reactance $X_n^{\scriptscriptstyle\rm TM}(ka)\!+\!\alpha_n X_n^{\scriptscriptstyle\rm TE}(ka)$ is zero at $k=k_0$ [and $R_n^{\scriptscriptstyle\rm TM}(k_0a)\!=\!R_n^{\scriptscriptstyle\rm TE}(k_0a)$ so that $R_n^{\scriptscriptstyle\rm TMTE}(k_0a)\!=\! 2R_n^{\scriptscriptstyle\rm TM}(k_0a)$].  With $Z_n^{\scriptscriptstyle\rm TM}(ka)$ from (\ref{ZTM}) and $Z_n^{\scriptscriptstyle\rm TE}(ka)$ from (\ref{ZTE}), we find
 \begin{subequations}
\lbl{ZTETM}
\be{ZTETMa}
 {Z_n^{\scriptscriptstyle\rm TMTE}(ka)=Z_n^{\scriptscriptstyle\rm TM}(ka)+\alpha_n Z_n^{\scriptscriptstyle\rm TE}(ka)}
\ee
\be{ZTETMb}
 {\alpha_n(k_0a) = |Z_n^{\scriptscriptstyle\rm TM}(k_0a)|^2/Z_f^2}.
\ee
\end{subequations}
Taking the frequency  derivative (that is, the derivative with respect to $ka$) of (\ref{ZTETMa}) and setting $k=k_0$, we get the self-tuned complex quality factor $\cQ^{\scriptscriptstyle\rm TMTE}_{nZ}(k_0a)$ as 
\begin{subequations}
\lbl{stTMTE}
\bea{stTMTEa}
\cQ^{\scriptscriptstyle\rm TMTE}_{nZ}(ka)= -jka Z_n^{\scriptscriptstyle\rm TMTE{\,\displaystyle\prime}}(ka)/[4R_n^{\scriptscriptstyle\rm TM}(ka)]\nonumber\\ = (1+\gamma_n) \widetilde{\cQ}^{\scriptscriptstyle\rm TM}_{nZ}(ka)/2\\[-7mm]\nonumber
\eea
 {\small\bea{stTMTEb}
 \widetilde{\cQ}^{\scriptscriptstyle\rm TM}_{nZ}(ka)\! = \!\frac{-jka Z_n^{\scriptscriptstyle\rm TM{\,\displaystyle\prime}}(ka)}{2R_n^{\scriptscriptstyle\rm TM}(ka)}\! =\! \frac{-jka Z_n^{\scriptscriptstyle\rm TE{\,\displaystyle\prime}}(ka)}{2\gamma_nR_n^{\scriptscriptstyle\rm TE}(ka)}\! =\! \frac{\widetilde{\cQ}^{\scriptscriptstyle\rm TE}_{nZ}(ka)}{\gamma_n(ka)}\\[-7mm]\nonumber
\eea}
\be{stTMTEc}
\gamma_n(ka) \!=\! -\!\left(|Z_n^{\scriptscriptstyle\rm TM}|/Z_n^{\scriptscriptstyle\rm TM}\right)^2
\! =\! -\!\left(Z_n^{\scriptscriptstyle\rm TE}/|Z_n^{\scriptscriptstyle\rm TE}|\right)^2\!,\;|\gamma_n|\!=\! 1
\ee
\end{subequations}
\normalsize
with $\widetilde{\cQ}^{\scriptscriptstyle\rm TM}_{nZ}(ka)$ the complex quality factor of the \textit{untuned} TM spherical mode [which is not equal to $\widetilde{\cQ}^{\scriptscriptstyle\rm TE}_{nZ}(ka)$ even though their magnitudes are equal, as shown in ({\ref{QZunt})].  In (\ref{stTMTE}), the tuned frequency $k_0$ has been relabeled as $k$.  The $Q^{\scriptscriptstyle\rm TMTE}_{nZ}(ka)=|\cQ^{\scriptscriptstyle\rm TMTE}_{nZ}(ka)|$ are nearly equal to $Q_{n\rm Chu}(ka)/2$ for $ka\la 1$ since $|1+\gamma_n|\approx 2$ and $\widetilde{Q}^{\scriptscriptstyle\rm TM}_{nZ}(ka)\approx Q_{n\rm Chu}(ka)/2$ for $ka\la 1$.
\par
With $\widetilde{Q}_{1Z}(ka)$ evaluated from (\ref{QZunt}) and $\gamma_n$ evaluated from (\ref{stTMTEc}), the $Q^{\scriptscriptstyle\rm TMTE}_{1Z}(ka)= |1+\gamma_n| \widetilde{Q}_{1Z}(ka)/2$ for self-tuned dipoles is given by
\bea{QZ1TME}
Q^{\scriptscriptstyle\rm TMTE}_{1Z}(ka)\!\!\!\!\!&=&\!\!\!\!\!\frac{1}{2(ka)^3}\frac{\sqrt{1\!+\!6(ka)^2 \!+\! 9(ka)^4 \!+\! 4(ka)^6}}{(1+ka^2)\sqrt{1 + (ka)^6}}\nonumber\\ \hspace{5mm}\!\!\!\!&\stackrel{ka\ll 1}{=}&\!\!\!\!\frac{1}{2}\left[\frac{1}{(ka)^3}+\frac{2}{ka}-ka\right] +O[(ka)^3]
\eea
and $Q^{\scriptscriptstyle\rm TMTE}_{1Z}(ka)$ decays as $1/(ka)^5$ for $ka\gg 1$.  This equation confirms that the minimum self-tuned dipole quality factor is equal to about half of the single-dipole quality factor in (\ref{QZ1TM}) or (\ref{QZ1TE}) for $ka\la 1$.
\subsection{Field-Based $Q_n(ka)$ for Spherical Modes}
Since the tuned TM or TE spherical-mode lower-bound quality factors in (\ref{QZnTM}) and (\ref{QZnTE}) are exactly equal to the field-base quality factors, they can also be obtained from (\ref{BQFs}) with \vspace{-5mm}
\bea{CWfsTMTE}
&&\hspace{-6mm} \cW_n \!= \frac{1}{4} \lim_{r\to \infty}\bigg[\!\int\limits_{\V_o(r>a)}\!\!\!\left[\mu_0|\vH_n|^2 +\eps_0|\vE_n|^2\right]d\V\nonumber\\
 &&\hspace{-10mm}- 2\eps_0 r \!\int\limits_{4\pi}|\vF_n|^2\, d\Omega \bigg]
 - \frac{j\eps_0 a}{2}\int\limits_{4\pi}\vF_n'\cdot\vF_n^*\, d\Omega + W_n^s
\eea
where the fields in (\ref{CWfsTMTE}) are those of the TM or TE spherical modes that produce the impedance $Z_n^{\scriptscriptstyle\rm TM}$ or $Z_n^{\scriptscriptstyle\rm TE}$, respectively, $W_n^s$ is the energy stored in the series tuning element, and the prime on $\vF_n'$ is now the derivative with respect to $ka$.  The corresponding energy formula used by Collin and Rothschild to get the tuned TM or TE quality factors (that are equal to the tuned TM or TE quality factors of Chu) is
{ \bea{CRQ}
 W_{n\rm CR} = \frac{1}{4} \lim_{r\to \infty}\bigg[\!\int\limits_{\V_o(r>a)}\!\!\left[\mu_0|\vH_n|^2 +\eps_0|\vE_n|^2\right]d\V\hspace{1mm}\nonumber\\ - 2\eps_0 (r-a) \int\limits_{4\pi}|\vF_n|^2\, d\Omega \bigg] +W_n^s.
\eea}\noindent
The difference between $\cW_n$ and $W_{n\rm CR}$ is
 \be{WWnCR}
 {\cW_n \!-\! W_{n\rm CR}\! =\! - \frac{\eps_0 a}{2} \int\limits_{4\pi}|\vF_n|^2\, d\Omega \!-\! \frac{j\eps_0 a}{2}\int\limits_{4\pi}\vF_n'\cdot\vF_n^*\, d\Omega}
\ee
which implies from (\ref{ZpWc})
{\small 
\bea{QQCR}
&&\hspace{-7mm}\cQ_n(ka)\! =\!  \frac{2\w}{|I|^2 R_n}\bigg[W_{n\rm CR}\! -\! \frac{\eps_0 a}{2}\! \int\limits_{4\pi}\!|\vF_n|^2\, d\Omega\! -\! \frac{j\eps_0 a}{2}\!\int\limits_{4\pi}\!\vF_n'\cdot\vF_n^*\, d\Omega\bigg]\nonumber\\
&&\hspace{12mm}=\bigg[Q_{n\rm CR}(ka) - ka - jka\frac{\int_{4\pi}\vF_n'\cdot\vF_n^*\, d\Omega}{\int_{4\pi}|\vF_n|^2\, d\Omega}\bigg]
\eea}\noindent
where $Q_{n\rm CR}(ka)= Q_{n\rm Chu}(ka)$ and the $\vF_n'\cdot\vF_n^*$ integral is generally complex valued.  The $\cQ_n(ka)$ given by (\ref{QQCR}) will in general be different for series tuned TM and TE spherical modes because the $\vF_n'\cdot\vF_n^*$ integrals will have different values for the TM and TE spherical modes.
For electrically small minimum-Q spherical-mode antennas, the $\vF_n'\cdot\vF_n^*$ integral contributes negligibly to the quality factors for the value of $ka$ somewhat less than $n$.
\par
For a tuned antenna (reactance zero) $\cW_n$ in (\ref{CWfsTMTE}) can be rewritten as
{ \bea{Wnt}
&&\hspace{-6mm} \cW_n = \frac{1}{2}\lim_{r\to \infty}{\rm max}\bigg[\!\int\limits_{\V_o(r>a)}\!\!\!\!\!\!{\scriptsize\left(\!\!\begin{array}{c}\mu_0|\vH_n|^2 \\ \eps_0|\vE_n|^2\\ \end{array}\!\! \right)}d\V
 - \eps_0 r \int\limits_{4\pi}|\vF_n|^2\, d\Omega \bigg]\nonumber\\
&&\hspace{25mm} - \frac{j\eps_0 a}{2}\int\limits_{4\pi}\vF_n'\cdot\vF_n^*\, d\Omega
\eea}\noindent
where ``max'' denotes the larger of the top (TE mode) or bottom (TM mode) quantity within the square brackets.  For an untuned antenna ($W_n^s=0$), the $|\vH_n|^2$ and $|\vE_n|^2$ integrals have to be evaluated separately and added together.  This has been done for the Stratton \cite[sec. 7.11]{Stratton} spherical modes in the insightful article by Fante with $(r-a)$ as in (\ref{CRQ}) rather than $r$ as in (\ref{Wnt}) \cite{Fante-1969}.  Thus, the complex untuned field-based quality factors $\widetilde{\cQ}_n(ka)$ can be written as
\begin{subequations}
\lbl{untcQ}
\be{untcQa}
\widetilde{\cQ}_n(ka) = Q_{n\rm F}(ka)/2 - ka - jka\textstyle\frac{\int_{4\pi}\vF_n'\cdot\vF_n^*\, d\Omega}{\int_{4\pi}|\vF_n|^2\, d\Omega}
\ee
with
\bea{untcQb}
&&\hspace{-7mm}\frac{Q_{n\rm F}(ka)}{2} = \w\bigg\{\frac{1}{4} \lim_{r\to \infty}\bigg[\!\int\limits_{\V_o(r>a)}\!\!\left[\mu_0|\vH_n|^2 +\eps_0|\vE_n|^2\right]d\V\hspace{1mm}\nonumber\\&& - 2\eps_0 (r-a) \int\limits_{4\pi}|\vF_n|^2\, d\Omega \bigg]\bigg\}\bigg/\textstyle\frac{1}{2Z_f}\!\!\int\limits_{4\pi}|\vF_n|^2\, d\Omega
\eea
\end{subequations}
where $Q_{n\rm F}(ka)$ are the Fante quality factors \cite{Fante-1969} (which are identical for TM and TE spherical modes).  The magnitudes, $\widetilde{Q}_n(ka)=|\widetilde{\cQ}_n(ka)|$ of the untuned spherical modes are equal to $\widetilde{Q}_{nZ}(ka)$ in (\ref{QZunt}), which are the same for the TM and TE spherical modes, even though $\widetilde{\cQ}^{\rm\scriptscriptstyle TM}_n(ka) \neq \widetilde{\cQ}^{\rm\scriptscriptstyle TE}_n(ka)$.
\par
{\color{black}For self-tuned TM-plus-TE spherical modes, $W_n^s$ in (\ref{CWfsTMTE}) is zero and the $\vE_n$ and $\vH_n$ fields are the sum of the electric and magnetic fields of both modes such that these fields produce the impedance $Z_n^{\scriptscriptstyle\rm TMTE}(ka)$ in (\ref{ZTETMa}) when they are integrated in the complex Poynting theorem.  However, only the TM fields are required if we note that $\cQ^{\rm\scriptscriptstyle TMTE}_n(ka)=\cQ^{\rm\scriptscriptstyle TMTE}_{nZ}(ka)$ and $\widetilde{\cQ}^{\rm\scriptscriptstyle TM}_n(ka)=\widetilde{\cQ}^{\rm\scriptscriptstyle TM}_{nZ}(ka)$, so that (\ref{stTMTEa}) and (\ref{untcQa}) give
\bea{QQCRME}
\cQ^{\scriptscriptstyle\rm TMTE}_n(ka)= \frac{1}{2}(1+\gamma_n) \widetilde{\cQ}^{\scriptscriptstyle\rm TM}_n(ka)
= \frac{1}{2}(1+\gamma_n)\nonumber\\ \cdot\bigg[Q_{n\rm F}(ka)/2 - ka - jka\textstyle\frac{\int_{4\pi}\vF_n^{TM\prime}\cdot\vF_n^{TM*}\, d\Omega}{\int_{4\pi}|\vF^{TM}_n|^2\, d\Omega}\bigg].
\eea
To confirm that the right-hand side of (\ref{QQCRME}) does indeed equal $\cQ^{\scriptscriptstyle\rm TMTE}_{nZ}(ka)$ in (\ref{stTMTEa}), the ratio of the far-electric-field integrals in (\ref{QQCRME}) must be evaluated.  This requires the $\vE_n^{\scriptscriptstyle\rm TM}$ field of the TM spherical mode that produces Chu's input impedance $Z_n^{\scriptscriptstyle\rm TM}(ka)$ of the TM spherical mode, in order to maintain the equality between the impedance-based and field-based quality factors [$\cQ(ka) = \cQ_Z(ka)$].  Applying the complex Poynting theorem to the volume enclosed by the surface of the shielded power supply plus reference plane of a hypothetical waveguide feeding (from inside the sphere) the equivalent currents producing the TM spherical mode fields, and the spherical free-space surface at $r=a$, reveals that in order to obtain the input impedance $Z_n^{\scriptscriptstyle\rm TM}(ka)$, the far electric field of the TM spherical mode has to satisfy the proportionality $\vF_n^{\scriptscriptstyle\rm TM} \propto Z_f I\vthh/[ka h_n^{(2)}(ka)]$.  This proportionality can be achieved without changing the value of the Stratton \cite[ch. 7]{Stratton} TM modal coefficients $b_n$ at each frequency $\w_0$ (that is, $k_0 a$) at which the complex Q-energy $\cW_n$ is evaluated by simply multiplying $b_n$ by\\[-3mm] 
\be{TMprop}
 { {\rm f}_n^{\scriptscriptstyle\rm TM} = \frac{k_0a h_n^{(2)}(k_0a)}{ka h_n^{(2)}(ka)}\frac{k}{k_0}}.
\ee
\textit{In other words, Chu's physically realizable (finite number of frequency independent resistors, inductors, and capacitors) TM spherical-mode input impedances implicitly require the corresponding Stratton  TM spherical-mode coefficients to be renormalized (and, similarly, the counterpart renormalization in (\ref{TEprop}) for TE spherical-mode coefficients).}   With this ${\rm f}_n^{\scriptscriptstyle\rm TM}$ factor in $\vF_n^{\scriptscriptstyle\rm TM}$, the ratio of the far electric field integrals in (\ref{QQCRME}) is equal to simply $jZ_n^{\scriptscriptstyle\rm TM}(ka)/Z_f = -(ka h_n^{(2)})'/(ka h_n^{(2)})$ and $\cQ^{\scriptscriptstyle\rm TMTE}_n(ka)$ becomes
\bea{QQCRME'}
\cQ^{\scriptscriptstyle\rm TMTE}_n(ka)= \frac{1}{2}(1+\gamma_n) \widetilde{\cQ}^{\scriptscriptstyle\rm TM}_n(ka)
= \frac{1}{2}(1+\gamma_n)\nonumber\\ \cdot\big[Q_{n\rm F}(ka)/2 - ka + jka\, (ka h_n^{(2)})'/(ka h_n^{(2)})\big]
\eea
where $\widetilde{\cQ}^{\scriptscriptstyle\rm TM}_n(ka)$ can be found from the second equation in (\ref{QQCRME'}).  From (\ref{QQCRME'}) and (\ref{Wnt}), $\cQ^{\scriptscriptstyle\rm TM}_n(ka)= Q_{nChu}(ka)-ka+jka[kah_n^{(2)}(ka)]'/[ka h_n^{(2)}(ka)]$.
Computations of $\cQ^{\scriptscriptstyle\rm TMTE}_n(ka)$ in (\ref{QQCRME'}) and $\cQ^{\scriptscriptstyle\rm TMTE}_{nZ}(ka)$ from the first equation in (\ref{stTMTEa}) confirms that they are equal.
\par
For the sake of completeness, a similar analysis with the Chu TE input impedances reveals that the far electric fields of the Stratton TE spherical-mode coefficients have to be renormalized by\\[-3mm]
\be{TEprop}
 { {\rm f}_n^{\scriptscriptstyle\rm TE} = \frac{[k_0a h_n^{(2)}(k_0a)]'}{[ka h_n^{(2)}(ka)]'}\frac{k}{k_0}}
\ee
so that (\ref{untcQa}) gives
 \be{untcQaTE}
 {\widetilde{\cQ}^{\rm\scriptscriptstyle TE}_n(ka)\! =\! Q_{n\rm F}(ka)/2 \!-\! ka \!+\! jka\, (ka h_n^{(2)})''/(ka h_n^{(2)})'}.
\ee
From (\ref{untcQaTE}) and (\ref{Wnt}), $\cQ^{\scriptscriptstyle\rm TE}_n(ka)= Q_{nChu}(ka)-ka+jka[kah_n^{(2)}(ka)]''/[ka h_n^{(2)}(ka)]'$.
Computations of $\widetilde{\cQ}^{\scriptscriptstyle\rm TE}_n(ka)$ in (\ref{untcQaTE}) and $\widetilde{\cQ}^{\scriptscriptstyle\rm TM}_n(ka)$ from (\ref{QQCRME'}) confirms that they satisfy the relationship obtained from (\ref{stTMTEb}), namely $\widetilde{\cQ}^{\scriptscriptstyle\rm TM}_n(ka)\! =\! \widetilde{\cQ}^{\scriptscriptstyle\rm TE}_n(ka)/\gamma_n$.  The factors ${\rm f}_n^{\scriptscriptstyle\rm TM}$ and ${\rm f}_n^{\scriptscriptstyle\rm TE}$ contribute only to $\vF'_n$ in the complex Q-energy $\cW_n$.
%\par
%\textit{Notwithstanding the renormalizations of the Stratton spherical modes in (\ref{TMprop}) and (\ref{TEprop}), it is emphasized that at each tuned frequency $\w_0$ (that is, $k_0a$) the spherical modes are multiplied by constants that keep the power in the spherical modes the same as before the renormalizations.}
%
\section{Maximum Gain and Supergain\lbl{MGS}}
Harrington \cite[sec. 6-13]{Harrington} has proven that for an antenna with  electric (TM) and magnetic (TE)  vector spherical modes up to degree $n\! = \!N$, the maximum possible gain of the antenna is
\be{Geq}
  {G = N(N+2),\;\; N\ge 1}.
\ee
This maximum gain occurs for self-tuned, cophasal, cross-polarized, $m\!=\!1$ spherical TM and TE multipoles with equal radiated power.
[For electric or magnetic spherical modes alone, the maximum possible gain is half the value in (\ref{Geq}).]
Thus, the gain of an antenna with its sources contained within a finite circumscribing radius $a_0$ can, in principle, be made arbitrarily large provided spherical modes are allowed with arbitrarily large $n\!=\!N$.  The recent result for the radius $a$ of the significant reactive power of an antenna with a maximum degree $n\!=\!N$ of significant spherical modes in the far field \cite{Yagh-Gen-2024}, namely\footnote{Remarkably, in regard to determining its far-field distance (generalized Rayleigh distance $R = 8a^2/\lambda$), the effective radius of an antenna is simply the radius $a$ to which the significant reactive power of the antenna extends \cite{Yagh-Gen-2024}, \cite{sampling}.}
\mbox{}\\[-3mm]
\be{conv}
 {a = \frac{N+1/2}{k},\;\; N\ge 1}
\ee
implies that such $G=N(N+2)$ high-gain antennas will also have high reactive power extending to a radius $a \approx N/k$, so that as the gain approaches an infinite value ($N\to\infty$), high reactive power will extend to an infinite radius ($a\to\infty$) from the center of the antenna whose sources are located within the finite radius $r=a_0$.  (The total reactive power and quality factor $Q$ of this antenna will also approach $\infty$ as $N\to\infty$.)
\par
The maximum gain formula in (\ref{Geq}) can be written in terms of $ka$, the electrical size of the significant reactive power, by substituting $N = ka-1/2$ from (\ref{conv}) into (\ref{Geq}) to get
\bea{Geq'}
  {G \!=\! (ka\!-\!1/2)(ka\!+\!3/2)\! =\! (ka)^2\! +\! ka\! -\! 3/4,\; ka\ge 3/2}
\eea 
which implies that $G\ge 3$, where $G=3$ is the maximum gain of electric-plus-magnetic dipoles ($N=1, ka=3/2$).
\par
Ordinary (nonsuper-reactive) antennas can be defined as having large reactive power extending no further than a radius equal to about $a = a_0 +3\lambda/(4\pi)$, that is, about a radial distance $\lambda/4$ beyond the sources of the antenna.\footnote{The value of $a = a_0 +3\lambda/(4\pi)$ inserted into (\ref{conv}) gives the degree number of spherical modes as $N_0 \approx [ka_0]+1$ (a result also gotten in \cite{Pigeon}) for obtaining reasonably accurate far fields of ``ordinary'' antennas. %(about $1\%$ or less mainbeam error for any $ka_0$).
For electrically large antennas ($ka_0\gg 1$), this $N_0$ is smaller than the $ka_0$-dependent degree number of spherical modes found by Song and Chew \cite[sec. 3.4.1]{Song}, namely $N^{\scriptscriptstyle\rm SC}_0 \approx ka_0 + [\ln(1/\varepsilon)]{\scriptscriptstyle^{2/3}}(ka_0)\scriptscriptstyle^{1/3}$, where $\varepsilon$ is the maximum fractional error in the far fields.  This expression was derived from the approximate value of the last term in the truncated spherical-mode expansion of an acoustic monopole displaced a distance $a_0$ from the origin.  However, the displaced monopole is not an ``ordinary'' antenna in that the high reactive power of this ideal point-source at $r=a_0$ gives infinite reactive energy and $Q$ (because of the inverse square singularity of the acoustic monopole velocity), unlike the finite reactive energy and low $Q$ of realistic electrically large antennas such as horn or horn-fed reflector antennas.  %For example, numerical evaluation of the truncation errors for a circular piston (ordinary) radiator \cite{W&Y} reveals that the value of $N_0 \approx ka_0+1$ gives a maximum mainbeam far-field error of about $-20$ dB ($1\%$) or less for any value of $ka_0$, and an error in total radiated power of about $-35$ dB or less.  A $ka_0 =1000$ gives a maximum mainbeam far-field error of about $-30$ dB with less than $\pm 1$ dB error in the lowest sidelobes, which are more than $80$ dB below the mainbeam.  Further numerical experimentation with the piston radiator shows that $N_0\approx ka_0+n_0$ with the integer $n_0>1$ gives a maximum mainbeam far-field error appreciably less than $-20$ dB for any value of $ka_0$; the maximum error decreasing with increasing values of $n_0$.  For example, $n_0=5$ gives a maximum mainbeam far-field error of about $-30$ dB ($0.1\%$) for any value of $ka_0$.  
The inapplicability of $N^{\scriptscriptstyle\rm SC}_0$ to ``ordinary'' antennas, as opposed to $N_0$, was not pointed out in \cite{Yagh-Gen-2024}.}   Consequently, the maximum gain $G_0$ of ordinary electric plus magnetic multipole antennas is given from (\ref{Geq'}) as
\be{Gmax}
 { G_0 = (ka_0+1)(ka_0+3) = (ka_0)^2 +4ka_0 +3}.
\ee
Thus, it may seem reasonable to define a supergain antenna as having a gain
 \be{sgain}
 {G > G_0 = (ka_0)^2 +4ka_0 +3}
\ee
where $a_0$ is the radius of the smallest sphere that circumscribes the sources of the antenna. 
However, it is found in engineering practice to be very challenging to obtain gains greater than that of a maximum-gain Huygens source (elementary electric and magnetic dipoles with a gain of $3$) for an antenna with electrical size $ka_0\la 1$ \cite{super}.  In other words, although (\ref{sgain}) is  a valid theoretical criterion for defining supergain based on the electrical size of the significant reactive-power radius of ``ordinary antennas'', in the transition region between $ka_0$ electrically small and $ka_0$ electrically large (say $0.5 \la ka_0\la \pi$), it is a difficult engineering challenge to design the quadrupoles and higher order multipoles in a small enough region of space to obtain the theoretically possible ordinary (nonsuper-reactive) gain in (\ref{Gmax}).   Therefore, it is more realistic to define supergain as \cite{Yagh-Gen-2024}
 \be{sgreal}
 {G(ka_)) > \bigg\{\begin{array}{ll} 3+ka_0\,,\hspace{17.5mm} ka_0\le 1\\(ka_0)^2 +ka_0+2\,,\quad ka_0\ge 1   \end{array}}
\ee
which is comparable to the supergain criterion, $G(ka_0) > (ka_0)^2 +3$, proposed by Kildal and Best \cite{K&B}.
For electrically large antennas ($ka_0 \gg 1$), the supergain criteria in both (\ref{sgain}) and (\ref{sgreal}), and that of Kildal and Best in \cite{K&B}, reduce to simply
\be{sgain1}
G(ka_0) \ga (ka_0)^2.
\ee
\par
{\color{black}The on-axis gain of a circular disk antenna of radius $a_0$ with uniform orthogonal electric and magnetic surface currents (Huygens currents) on the disk is the same as the gain of a plane wave illuminating a uniform circular aperture \cite[ch. 10]{Johnson}, \cite{Neto}.  Specifically, this on-axis gain $G_{\rm dk}(ka_0)$ is found to be
 \be{oag}
 G_{\rm dk}(ka_0) = \frac{(ka_0)^2}{\left[1 - \frac{1}{2ka_0}\int_0^{2ka_0} J_0(u)du\right]}
\ee
where $J_0(u)$ is the zeroeth order Bessel function.
The denominator in (\ref{oag}) is not present in \cite[ch. 10]{Johnson} because in finding the gain of the circular aperture antenna, Johnson uses the plane-wave power in the aperture rather than the exact power radiated by the far-field pattern of the disk used to get (\ref{oag}).  %, where it can be shown that\\[-3mm]
%\be{oag'}
% {\int\limits_0^{\pi/2} \!\frac{1+\cos^2{\theta}}{\sin{\theta}} J_1^2(ka_0 \sin{\theta}) d\theta = 1 - \frac{1}{2ka_0}\!\int\limits_0^{2ka_0}\! J_0(u)du}.\\[-1.5mm]
%\ee
\par
The gain of the Huygens current disk is the same as a hypothetical uniform-aperture reflector antenna, which can be considered an ``ordinary'' (nonsuper-reactive) self-tuned antenna.\footnote{Uniform-surface-current disk antennas have infinite reactive energy and $Q$ because of the line charges at the rim of the disk caused by the abrupt discontinuity in the surface current.  This line charge and infinite reactive energy can be eliminated with an edge current that satisfies the finite-energy condition \cite[sec. 9.7.5]{vanBladel} near the rim of the disk without appreciably changing the gain.} %with well-behaved equivalent surface currents and significant reactive power extending no further than about $\lambda/(2\pi)$ from the rim of the disk.  
Its gain should be reasonably close to but less than the lower limit for supergain given in (\ref{sgreal}) because its far-field pattern has about $N = ka_0$ degree significant spherical modes \cite{W&Y} and a gain of about $(ka_0)^2$ for $ka_0\gg 1$, very close to the maximum gain in (\ref{Geq}) for $N=ka_0\gg 1$.  Also, $G_{\rm dk}(ka_0)$ in (\ref{oag}) approaches the correct electric-plus-magnetic Huygens dipole value of $3$ as $ka_0\to 0$.  Plots of the right-hand side of (\ref{sgreal}) and $G_{\rm dk}(ka_0)$ from (\ref{oag}) vs $ka_0$ in Fig. \ref{Gfig} indicates that indeed the lower limit given in (\ref{sgreal}) is a realistic criterion for defining supergain.  The supergain curve of Kildal and Best \cite[$(ka_0)^2 + 3$]{K&B}  is shown by the dash-dot line in Fig. \ref{Gfig}.  This [$(ka_0)^2 + 3$] supergain curve falls below the supergain curve from (\ref{sgreal}) and below the circular-disk gain curve for $ka_0 \ga 5$;  thus the Kildal and Best curve is a relatively lenient supergain criterion.}
\begin{figure}[h]
\vspace{-4mm}
\hspace{-4mm}
\includegraphics[height=1.8in,width=3.8in]{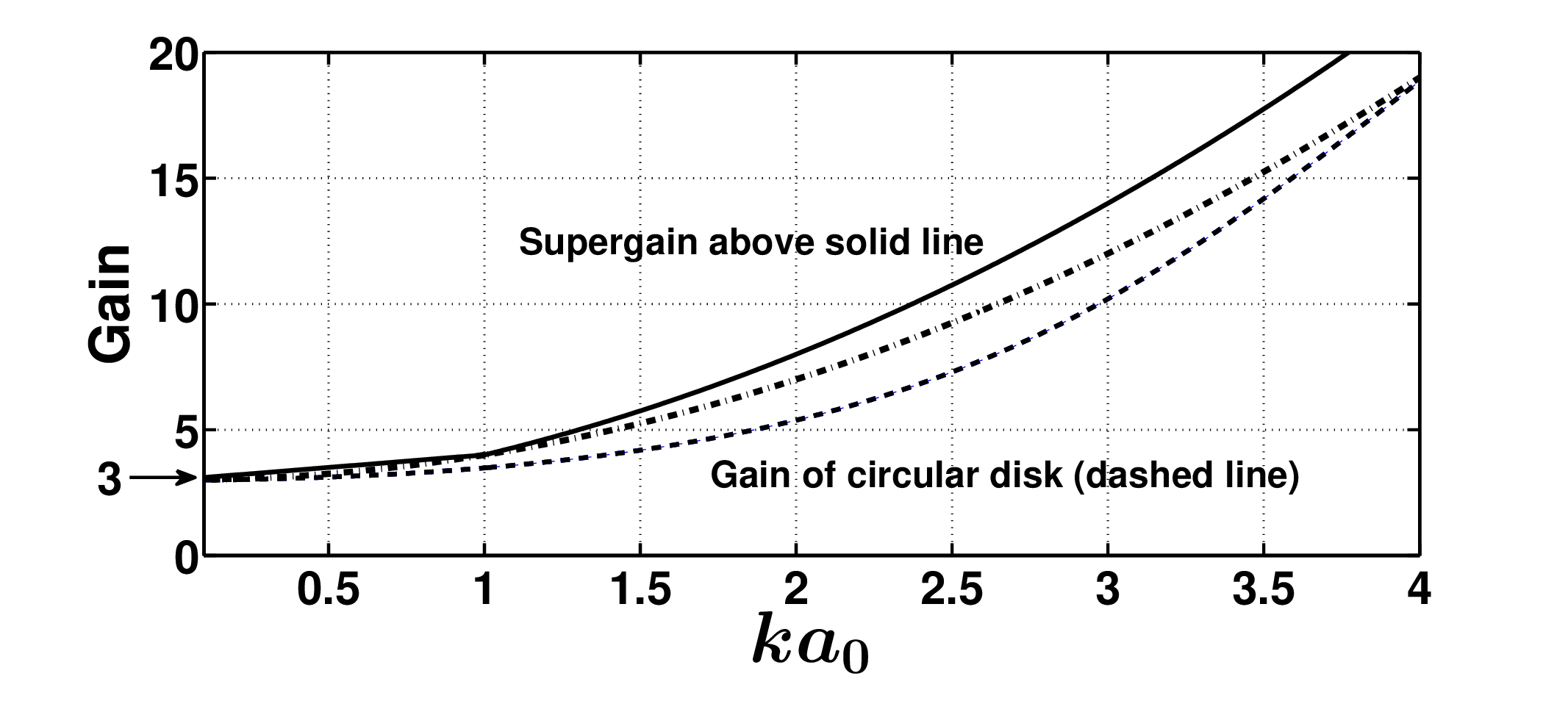}
\vspace{-7mm}
\caption{Gain $G_{\rm dk}$ vs $ka_0$ (dash line) of the circular disk with radius $a_0$ of uniform electric and magnetic surface currents (high-gain ``ordinary'' antenna) compared to the supergain lower limit vs $ka_0$ on the right-hand side of (\ref{sgreal}) (solid line) and that of Kildal and Best \cite{K&B} (dash-dot line).}
\label{Gfig}
%\mbox{}\\[-11mm]
\end{figure}
\par
Predicted and measured supergains of about $G=5$ (7 dB) were first obtained by Yaghjian et al. \cite{super, Yaghjian-2005,O&Y},  and later by Lim and Ling \cite{L&L}, for two-element supergain endfire arrays that fit inside spheres with electrical sizes $ka_0$ between 0.5 and 1.0.  The $ka_0 \approx 1$ two-element $7$-dB supergain endfire array in \cite{super} has a center frequency of about $0.875$ GHz with $98.5$\% efficiency and a quality factor of $41$ ($\approx 5$\% half-power VSWR bandwidth) matched to $50$ ohms.
\subsection{\lbl{MG}Maximum Gain Versus Minimum Quality Factor for Spherical Antennas with Degree $N$ Spherical Modes}
Harrington \cite[sec. 6-13]{Harrington} also determined the minimum quality factor for a spherical antenna of radius $a$ that produces the $N$-degree spherical modes with the maximum gain in (\ref{Geq}).  (Here, and throughout the rest of the paper, $a$ denotes the radius of the physical antenna, not the radius of the significant reactive fields, and again the term ``minimum'' means that there is no contribution to the quality factor from inside the sphere of radius $a$, only the contribution from the fields outside the sphere.)   Specifically, he finds in \cite[eq. 6-162]{Harrington} the following expression for the minimum quality factor of a maximum-gain $N$-degree self-tuned spherical antenna of radius $a$
 \be{HGQ}
 {Q_{\rm H}(ka) = \textstyle{\frac{1}{2N(N+2)}}\sum\limits_{n=1}^N (2n+1) Q_{n\rm Chu}(ka)}
\ee
where the $Q_{n\rm Chu}(ka)=Q_{n\rm CR}(ka)$ are the Chu/CR spherical-mode quality factors.  No tuning capacitor or inductor is required because the TM and TE modes are assumed present with equal power radiated and thus equal electric-field and magnetic-field energy giving zero reactance.\footnote{\lbl{thal}{\color{black}As mentioned in Footnote \ref{CRfoot}, tuning elements are required to obtain the required phase difference between the TM and TE spherical modes of electrically small antennas that could add to the value of the quality factors \cite{Thal-2009}.  However, dispersive tuning could avoid this increase in quality factor without changing the directivity but with a reduced gain and bandwidth power drop; see Section \ref{DT}.  Without dispersive tuning, the $ka = 0.2, 0.5, 1.0$ curves in Fig. \ref{GvQ} would shift at the bottom ($G=3$) to the right by a $\log_{10} Q$ amount equal to about $0.3, 0.2, 0.06$, respectively, and at the top ($G=8$) by about $0.3, 0.27, 0.19$ \cite[tables I and II]{Thal-2009}.  With dispersive tuning that reduces the antenna efficiency to $50\%$, both the gain and quality factor would halve.  In other words, the horizontal and vertical scales in Fig. \ref{GvQ} would change from $(-0.2,3)$ and $(3,8)$ to $(-0.5,2.7)$ and $(1.5,4)$, respectively.}}  The Stratton \cite[ch. 7]{Stratton} TE and TM spherical mode coefficients ($a_n,b_n$) are equal for maximum gain antennas such that $a_n/a_1 = b_n/b_1 = (2n+1)/[n(n+1)]$.  [In his computations, Harrington actually uses the approximation to $Q_{n\rm Chu}(ka)$ that Chu uses in his paper rather than the Chu ladder-network quality factors that are denoted herein by $Q_{n\rm Chu}(ka)$.]  Because the TM and TE modes in (\ref{HGQ}) are self-tuned and require no tuning inductor or capacitor, the spherical quality factors of Chu in (\ref{HGQ}) ignore the electric-field energy of the TE modes and the magnetic-field energy of the TM modes.  Fante \cite{Fante-1969} corrects this deficiency by adding a $Q_n^{\rm p}(ka)$ (Fante uses a superscript prime rather than the p here) to $Q_{n\rm Chu}(ka)$ in (\ref{HGQ}) to get
 \be{HGQF}
 {Q_{\rm F}(ka) = \textstyle{\frac{1}{2N(N+2)}}\sum\limits_{n=1}^N (2n+1)Q_{n\rm F}(ka)}
\ee
where the spherical-mode Fante quality factors are defined here as
 \begin{subequations}
\lbl{Fqf}
\be{Fqfa}
 {Q_{n\rm F}(ka) = Q_{n\rm Chu}(ka) + Q_n^{\rm p}(ka)}
\ee
which are also given in (\ref{untcQb}) in terms of the fields of the self-tuned TM-plus-TE spherical modes.
For $n=1$,  $Q_1^{\rm p}(ka) = 1/(ka)$, so that (\ref{QChu}) and (\ref{Fqfa}) give
\be{Fqfb}
 {Q_{1\rm F}(ka) = \frac{1}{(ka)^3} + \frac{2}{ka}}.
\ee
\end{subequations}
One half the value of this $Q_{1\rm F}(ka)$ equals the Fante quality factor for a self-tuned spherical TM-plus-TE dipole \cite{Fante-1969}, \cite{McLean}, that is, $Q_{1\rm F}^{\scriptscriptstyle\rm TMTE}(ka) = [1/(ka)^3 + 2/ka]/2$, which agrees with the first two terms of $Q_{1Z}^{\scriptscriptstyle\rm TMTE}(ka)$ in (\ref{QZ1TME}).}
The paper by Passalacqua et al. \cite{Passalacqua-2023} on ``Q-bounded maximum directivity'' uses the Fante quality factors as given in (\ref{Fqfa}).
\par
However, we now know that one can obtain a more accurate value than either $Q_{\rm H}(ka)$ in (\ref{HGQ}) or $Q_{\rm F}(ka)$ in (\ref{HGQF}) by using the most accurate possible expressions for the complex quality factors of the self-tuned spherical modes, namely the $\cQ^{\scriptscriptstyle\rm TMTE}_n(ka)=\cQ^{\scriptscriptstyle\rm TMTE}_{nZ}(ka)$ in (\ref{stTMTEa}) and (\ref{QQCRME}).  {\color{black}Then repeating Harrington's derivation for the minimum quality factor of an antenna with the maximum gain \cite[sec. 6-13]{Harrington},  but now with the self-tuned spherical-mode complex quality factors $\cQ^{\scriptscriptstyle\rm TMTE}_{nZ}(ka)$ in (\ref{stTMTEa}), and noting from (\ref{CWfsTMTE}) with $W_n^s=0$ that the complex $\cQ^{\scriptscriptstyle\rm TMTE}_n(ka)$ [which are equal to $\cQ^{\scriptscriptstyle\rm TMTE}_{nZ}(ka)$] maintain orthogonality, one gets instead of (\ref{HGQ}) or  (\ref{HGQF})
 \be{HGQA}
 {Q_{\rm A}(ka) = \textstyle{\frac{1}{2N(N+2)}}\left|\sum\limits_{n=1}^N (2n+1)(1 + \gamma_n)\widetilde{\cQ}^{\scriptscriptstyle\rm TM}_{nZ}(ka)\right|}
\ee
where from (\ref{stTMTEb})
 \be{Qnzut}
 {\widetilde{\cQ}^{\scriptscriptstyle\rm TM}_n(ka)\! =\! \widetilde{\cQ}^{\scriptscriptstyle\rm TM}_{nZ}(ka)\! = \!\text{-}j\frac{kaZ_n^{{\scriptscriptstyle\rm TM}\prime}(ka)}{2R^{\scriptscriptstyle\rm TM}_n(ka)}}
\ee
and, as explained above, $\gamma_n$ is the complex factor required to self-tune the TM and TE spherical modes so that their total reactance is zero and their radiated powers are the same.  The accurate $Q_{\rm A}(ka)$ can also be found directly from $\cQ^{\scriptscriptstyle\rm TMTE}_n(ka)$ in (\ref{QQCRME'}) as
\bea{HGQAfb}
 Q_{\rm A}(ka) = \textstyle{\frac{1}{2N(N+2)}}\Big|\sum\limits_{n=1}^N (2n+1)\hspace{25mm}\nonumber\\
\cdot(1+\gamma_n)\big[Q_{n\rm F}(ka)/2 - ka + ka\, Z_n^{\scriptscriptstyle\rm TM}(ka)/Z_f\big]\Big|.
\eea
The individual self-tuned TM-plus-TE complex quality factors are given by
\bea{TMpTE'}
&&\hspace{-6mm}\cQ^{\scriptscriptstyle\rm TMTE}_{nZ}(ka) \!=\! (1+\gamma_n)\widetilde{\cQ}_{nZ}(ka)/2\!=\! (1+\gamma_n)\widetilde{\cQ}_n(ka)/2\nonumber\\ && \hspace{-6mm} =(1+\gamma_n)\big[Q_{n\rm F}(ka)/2 - ka + ka\, Z_n^{\scriptscriptstyle\rm TM}(ka)/Z_f\big]\\ && \hspace{-6mm} = \!\cQ^{\scriptscriptstyle\rm TMTE}_n(ka)
 = [Q_{n\rm F}\!-\! ka(jZ_n^{{\scriptscriptstyle\rm TE}\prime}\!\!/Z_f \!-\! 2)Z_n^{\scriptscriptstyle\rm TM}\mskip-5mu/Z_f]/2\! -\! ka\nonumber
\eea
where the last equality is proven directly from $\cQ(\w) = \w \cW(\w)/P_{\rm\scriptscriptstyle R}$ and (\ref{CWfsTMTE}) using the [$\vE_n^{\scriptscriptstyle\rm TMTE},\vH_n^{\scriptscriptstyle\rm TMTE}$] fields with the factors ${\rm f}_n^{\scriptscriptstyle\rm TM}$ and ${\rm f}_n^{\scriptscriptstyle\rm TE}$ in (\ref{TMprop}) and (\ref{TEprop}) required to evaluate $\vF_n^{{\scriptscriptstyle\rm TMTE}\prime}(ka)$.  Computations confirm these equalities in (\ref{TMpTE'}).  The $|\cQ^{\scriptscriptstyle\rm TMTE}_{1Z}(ka)| = |\cQ^{\scriptscriptstyle\rm TMTE}_1(ka)|$ from (\ref{TMpTE'}) equal $Q^{\scriptscriptstyle\rm TMTE}_{1Z}(ka)$ in (\ref{QZ1TME}). 
\par
The maximum gain in (\ref{Geq}) for the $N$-degree self-tuned spherical-mode antenna of radius $a$ versus the minimum quality factors $Q_{\rm F}(ka)$ and $Q_{\rm A}(ka)$ in (\ref{HGQF}) and (\ref{HGQA}) is plotted in Fig. \ref{GvQ} for values of $ka = 0.2,\,0.5, \mbox{ and } 1.0$, without the extra tuning quality factors given in Footnote \ref{thal}.  The lines extrapolate between the discrete points which are at the finite degree-$N$ maximum gains from (\ref{Geq}).  The lines in Fig. \ref{GvQ} show that there is not much of a difference between $Q_{\rm F}(ka)$ and the more accurate $Q_{\rm A}(ka)$ except for $ka \ga 1$ with $Q$ on the order of unity or less where quality factor is usually not a practical measure of bandwidth.  The solid lines connecting the more accurate $Q_{\rm A}(ka)$ in Fig. \ref{GvQ} lie above the $Q_{\rm F}(ka)$ dashed lines. The smooth dash-dot maximum-gain versus minimum-quality-factor curves in Fig. \ref{GvQ} are obtained from Passalacqua et al. \cite[eqs. (9)-(10)]{Passalacqua-2023} (who allow an indefinitely large number of degree-$N$ spherical modes with the quality factors of Fante) but with the more accurate quality factors $|1 + \gamma_n| \widetilde{Q}_{nZ}(ka)$ from (\ref{HGQA}) replacing the Fante quality factors.\footnote{For $ka \la 1$, the absolute value signs in (\ref{HGQA}) can be brought inside the summation without making a significant change in the value of $Q_{\rm A}(ka)$.}  Although not shown in Fig. \ref{GvQ}, the $ka = 0.2$ dash-dot curve, like the $ka = 0.5$ and $1.0$ curves, swings to the left of the $Q_{\rm F}$ and $Q_{\rm A}$ circle markers at $G = 8$.   For $N = ka \gg 1$, $Q_{\rm F} \gg 1$ whereas $Q_{\rm A} < 1$, thereby revealing that $(Q_Z,Q_{\rm A})$ but not $(Q_{n\rm F},Q_{n\rm Chu},Q_{\rm F})$ give a valid measure of bandwidth for all $ka$.
\begin{figure}[h]
\vspace{-4mm}
\hspace{-5mm}
\includegraphics[height=1.8in,width=3.9in]{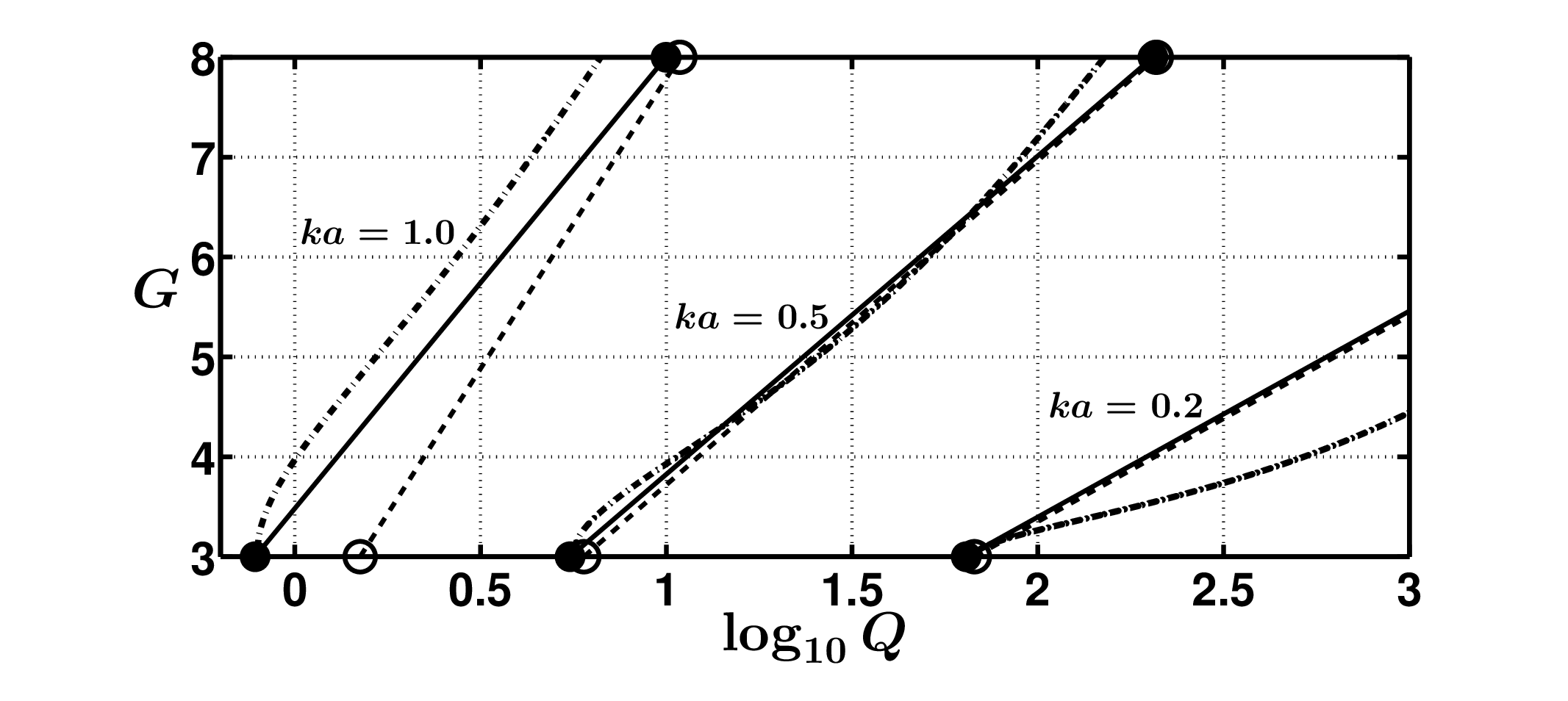}
\vspace{-7mm}
\caption{Maximum $N$-degree spherical-mode gain $G$ versus the minimum quality factor for the Fante quality factors $Q_{\rm F}(ka)$ (\boldmath$\scriptscriptstyle ---$\unboldmath)  and the more accurate $Q_{\rm A}(ka)$ quality factors (------) at the three different values of $ka = 0.2$, $0.5$, and $1.0$.  The lines extrapolate between the discrete points (circle markers) for $N = 1, 2$ ($G = 3, 8$).  The maximum-gain versus minimum-quality-factor curves obtained in \cite[eqs. (9)-(10)]{Passalacqua-2023} but with our more accurate quality factors in (\ref{HGQA}) are shown by the dash-dots ($-\!$ - $\!-$).}  %The squares mark the $[G_{\rm dk}(ka), Q_{\rm dk}(ka)]$ for the three different size Huygens current disks, $ka = 0.2$, $0.5$, and $1.0$.}
\label{GvQ}
%\mbox{}\\[-11mm]
\end{figure}
\par
According to (\ref{sgreal}), antennas with electrical sizes of $ka \la 1$ and gains larger than $3+ka$ in Fig. \ref{GvQ} should be classified as supergain antennas.  The $ka \approx 1$ antenna in \cite{super}, mentioned above in Section \ref{MGS}, with supergain equal to about $5$ ($7$ dB) had a quality factor of $41$.   The dash-dot $ka = 1$ line in Fig. \ref{GvQ}, combined with the extra tuning quality factors given in Footnote \ref{thal} for maximum gain self-tuned spherical modes, indicates that it is physically possible to have the same electrical-size and supergain antenna with a quality factor as low as about $3.3$.  For $ka = 0.5$, Fig. \ref{GvQ} with the extra tuning given in Footnote \ref{thal} indicates that this supergain of $5$ ($7$ dB) could be achieved with a quality factor as low as about $42$.
\section{Reducing the Antenna $Q$ with Dispersive Tuning\lbl{DT}}
Consider a one-port, feed-line matched, linear, passive, time-invariant, electrically small antenna (ESA--electrical size $ka \la 0.5$) tuned to a resonance or antiresonance at the angular frequency $\w$.  The field-based quality factor $Q(\w)$ of this antenna can be rewritten in (\ref{BQFf}) as
 \be{1}
 {Q(\w) = \eta(\w) \frac{\w |\cW(\w)|}{P_{\rm rad}(\w)}}
\ee
where $P_{\rm rad}(\w)$ is the average power radiated by the antenna and $\eta(\w) = P_{\rm rad}(\w)/P_{\rm\scriptscriptstyle R}(\w)$ is the radiation efficiency of the antenna that accepts a total average power of $P_{\rm\scriptscriptstyle R}(\w)$ (antenna-loss plus radiated average power).   The contribution to the complex Q-energy $\cW_{\!\left\{\substack{\rm e\\[-.25mm]\rm m}\right\}}(\w)$  from a series parallel-plate capacitor or solenoidal inductor used to tune the antenna to a resonance/antiresonance can be written as an integral over the volume $\V_{\rm e}$ or $\V_{\rm m}$ of the capacitor or inductor, respectively [see (\ref{CWfs}) with $\vF=0$, $\sigma_e =\w\eps_i$, $\sigma_m =\w\mu_i$] 
 \be{3}
 {\cW_{\!\left\{\substack{\rm e\\[-.25mm]\rm m}\right\}}(\w) = \frac{1}{4}\!\!\!\!\!\int\limits_{\quad\V_{\!{\scriptscriptstyle\left\{\substack{\rm\scriptscriptstyle e\\[-.25mm]\rm\scriptscriptstyle m}\right\}}}}\!\!\!\!\!\left\{\substack{(\w\mbox{ $\eps$})'(\eps^*/\eps) E^2\\ (\w\mbox{ $\mu$})' H^2}\right\} d\V}.
\ee 
It is assumed that the tuning capacitor or inductor is filled with material having scalar permittivity $\eps$ or scalar permeability $\mu$ that can be a complex function of $\w$, such that $E = -jI/(\w\eps A)$ in the capacitor and $H = {\cal N}I/d$ in the inductor, where $Ad$ is the volume of the parallel-plate capacitor or ${\cal N}$-turn solenoidal inductor. An alternative way to derive (\ref{3}) is to also use the impedances $Z =-jd/(\w\eps A)$ for the parallel-plate capacitor and $Z=j\w\mu {\cal N}^2 A/d$ for the solenoidal inductor in the relationship (\ref{ZpW}) for $\cW$ in terms of $Z'$.
\subsection{Nondispersive Tuning}
Before the publication of the references \cite{Y&G&J}, \cite{Y_AWPL}, \cite{Yaghjian-Reducing}--\cite{Yaghjian-IET}, lower bounds on $Q(\w)$ were determined  assuming implicitly that $\eps = \eps_e - j\sigma_e/\w$ and $\mu = \mu_m - j\sigma_m/\w$ with $\eps_e$, $\sigma_e$, $\mu_m$ and $\sigma_m$ in the tuning elements real-valued functions independent of frequency $\w$ (nondispersive) and position $\vrp$, so that (\ref{3}) reduces to the positive real stored energy \cite{Chu,C&R} 
 \be{4}
 {W_{\!\left\{\substack{\rm e\\[-.25mm]\rm m}\right\}}(\w) = \frac{1}{4}\Big\{\substack{\mbox{$\eps_e$} \\ \mbox{$\mu_m$}}\!\!\!\!\!\!\int\limits_{\quad\V_{\!{\scriptscriptstyle\left\{\substack{\rm\scriptscriptstyle e\\[-.25mm]\rm\scriptscriptstyle m}\right\}}}}\!\!\!\!\!\!\substack{ E^2\\ H^2}\Big\}\, d\V}.
\ee 
Note that $[\w(\sigma_{\!\left\{\substack{\rm\scriptscriptstyle e\\[-.25mm]\rm\scriptscriptstyle m}\right\}}/\w)]' = 0$.   For electrically small electric or magnetic dipoles that are not self-tuned, the untuned reactance is produced predominatly by the quasi-static electric or magnetic fields, respectively.  This implies that the tuning element must contribute an almost equal and opposite magnetic-field or electric-field reactance in the tuning inductor or capacitor, respectively, to make the total reactance $X(\w)=0$.  This, in turn, nearly doubles the Q-energy of the antenna through the addition of the energy in the tuning inductor or capacitor given in (\ref{4}).  Consequently, the minimum $Q(\w)$ can be found for these traditionally tuned electrically small magnetic or electric dipoles in a volume of arbitrary shape \cite{Y&G&J}.  For a spherical dipole antenna, this lower bound on $Q(\w)$ is given simply by the  Chu/CR lower bound for an electric or magnetic spherical dipole; see (\ref{QChu}), (\ref{QZ1TM}), and (\ref{QZ1TE}) as well as \cite{Chu,C&R}
\be{5}
 { Q_{1\rm Chu}(ka) = Q_{1\rm CR}(ka) = \frac{\eta(\w)}{(ka)^3},\quad ka \ll 1}
\ee 
with, as usual, $k=\w/c$ ($c$ the speed of light), $a$ is the radius of the sphere, and the efficiency $\eta(\w) < 1$ if there is loss.
\subsection{Dispersive Tuning}
The only linear, passive way to lower the single isolated resonance or antiresonance $Q(\w)$ in (\ref{1}) is to reduce the energy in the tuning element of the antenna below that of the frequency independent tuning-material elements assumed by Chu \cite{Chu}, Collin-Rothschild \cite{C&R}, and others following them.  This can be done with dispersive material in the tuning capacitor/inductor or by using a circuit that has the same impedance as the dispersive-material tuning element.  The idea of using a circuit equivalent to that of the dispersive-material tuning element was recently proposed and utilized by Radi and Al\`u \cite{Alu}.
\par
If the tuning of the electrically small capacitive or inductive antenna is done with an inductor or capacitor filled with frequency dispersive material such that $[\w\mu_i(\w)]'$ or $[\w\eps_i(\w)]'$ is negligible and 
 $[\w\mu_r(\w)]' <\mu_m(\w)$ or $[\w\eps_r(\w)]' <\eps_e(\w)$, respectively, where the subscripts $r$ and $i$ denote the real and imaginary parts, then the Q-energy of the antenna can be less than doubled and the lower bound on $Q(\w)$ is reduced with respect to that of the nondispersively tuned antenna.  In particular, it is shown in \cite{Y&G&J,Yaghjian-Reducing, Yaghjian-IET} that an initially  lossless magnetic-dipole antenna (for example) can be tuned with a Lorentzian permittivity having $[\w\eps(\w)]'=0$ in order to approximately halve the lower bound on $Q(\w)$ (twice the bandwidth) for about a $-25$ dB power-drop bandwidth and a 50\% efficiency, thus replacing the Chu/CR lower bound for a spherical electric or magnetic-dipole antenna in (\ref{5}) by
 \be{6}
 {Q(ka) = \frac{\eta(\w)}{2(ka)^3},\quad ka \ll 1}.
\ee 
In fact, as Al\`u has recently demonstrated \cite{Alu}, it is also possible to have a Lorentzian permittivity or permeability with $[\w\eps_r(\w)]' <0$ or $[\w\mu_r(\w)]'<0$, so as to reduce the lower bound on $Q(\w)$ further and thus increase the bandwidth by more than a factor of two (for an electric or magnetic di\-pole antenna with a small enough bandwidth power-drop and the same efficiency).
\par
The reactance and resistance shown in Fig. \ref{RX} \cite[fig. 2]{Yaghjian-Reducing} for the dispersively tuned antenna in Fig. \ref{fig1} \cite[fig. 1]{Yaghjian-Reducing} clearly shows that the dispersively tuned antenna maintains a single isolated resonance.  The only other way to linearly and passively increase the bandwidth beyond that predicted by the Chu/CR lower bound on Q is to add Bode-Fano tuning circuits to produce multiple resonances/antiresonances (as discussed in Section \ref{BFM}).  Presumably,  Bode-Fano tuning could be combined with dispersive tuning to further increase the bandwidth but at the expense of an increased ``group delay'' inherent in the multiple-resonance/antiresonance tuning.
\begin{figure}[h]
\vspace{-6.5mm}
\hspace{-3mm}
\includegraphics[height=1.8in,width=3.8in]{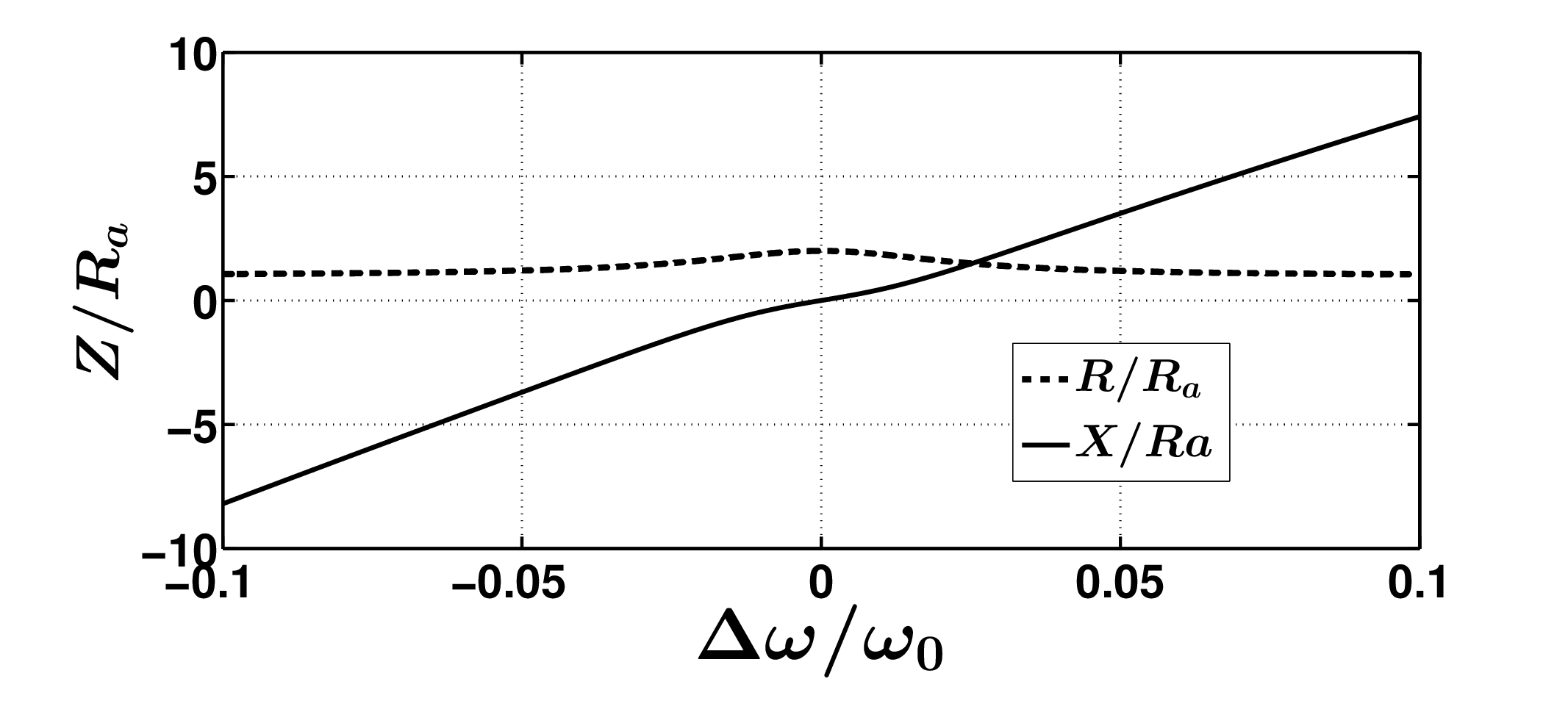}
\vspace{-7mm}
\caption{The input impedance for a dispersively tuned $Q = Q_Z = 10$ antenna with $50$\% efficiency and reduced fractional quality factor $\rho = 0.5$ (bandwidth doubled, that is, $Q_0 = Q_{0Z} = 20$ for the nondispersively tuned antenna with the same $50$\% efficiency). Note the single isolated resonance, which confirms that, unlike Bode-Fano tuning, the Lorentz dispersive tuning does not produce multiple resonances/antiresonances or increase the group delay.}
\label{RX}
\end{figure}
%\mbox{}\\[-6mm]
%
%
\subsection{Lorentzian Permittivity Tuning}
In \cite{Yaghjian-Reducing} and \cite{Yaghjian-IET}, the matched lossless inductive antenna having input inductance $L_a$ and radiation resistance $R_a$, as shown in Fig. \ref{fig1}, is tuned with a series capacitor filled with dispersive relative permittivity $\eps(\w)/\eps_0$ given by the passive (and thus causal \cite{Yagh-Passive}) Lorentzian function
\be{Y1}
 {\frac{\eps(\w)}{\eps_0}\! =\! \frac{\eps_{\rm r}(\w)\!-\!j\eps_{\rm i}(\w)}{\eps_0}\! =\! 1\! +\! \frac{\cal A}{1-(\w/\w_0)^2 + jg(\w/\w_0)}}
\ee
where $\w_0$ is the resonant frequency of the Lorentzian permittivity, and $\cal A$ and $g$ are adjustable positive real constants.  {\color{black}The purpose of the Lorentzian permittivity is not to introduce another resonance, as with Bode-Fano tuning, but to obtain an $\eps(\w)$ that allows $[\w\eps_i(\w)]' \approx 0$ and $[\w\eps_r(\w)]' <\eps_e(\w)$.}
\begin{figure}[h]
\vspace{-2mm}
%\hspace{-4mm}
\includegraphics[height=1.5in,width=3.6in]{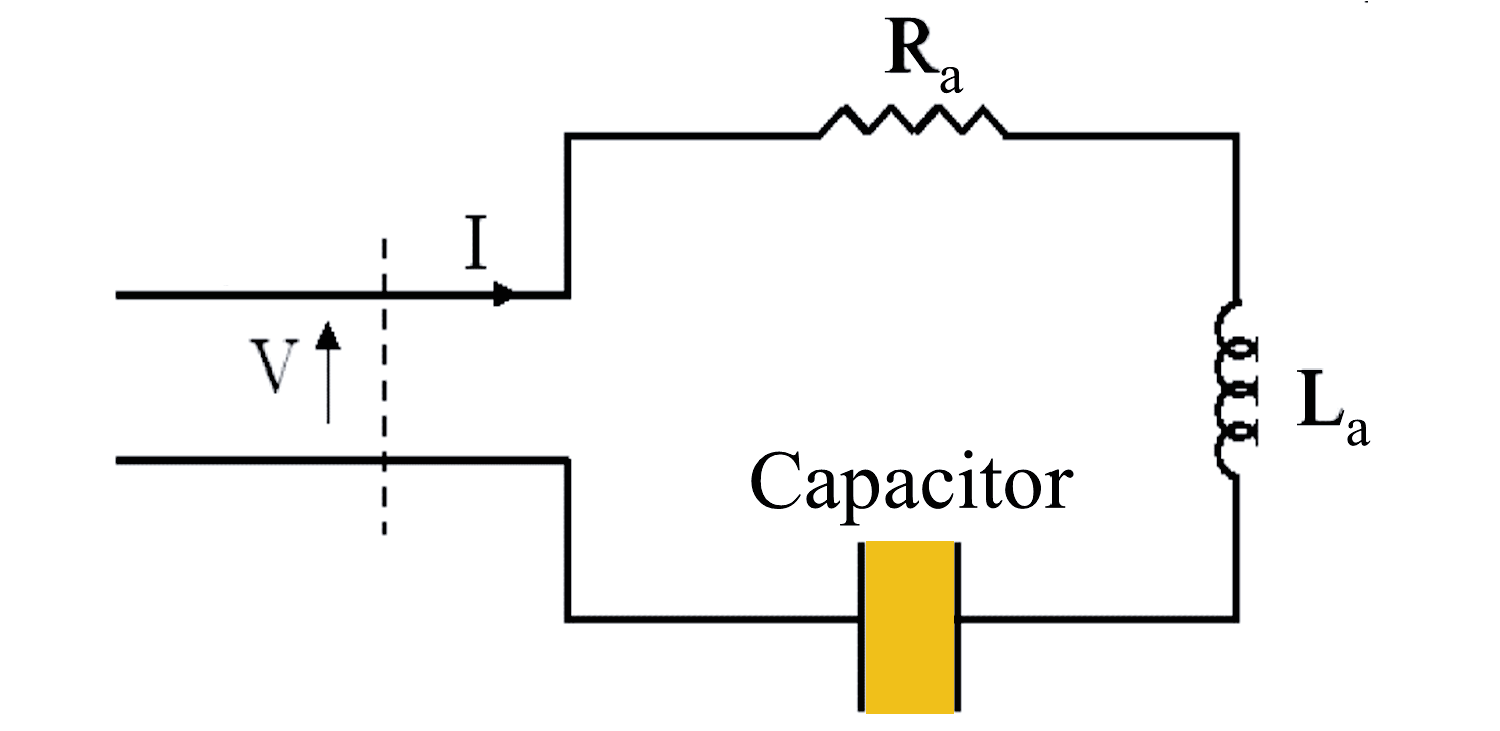}
\vspace{-7mm}
\caption{RLC series circuit of an inductive antenna tuned by a capacitor filled with a Lorentzian di\-elec\-tric material having relative permittivity given in (\ref{Y1}).}
\label{fig1}
%\mbox{}\\[-11mm]
\end{figure}
\par
In Fig. \ref{fig2}, taken from \cite{Yaghjian-Reducing}, the maximum power-drop factor $\alpha_{\rm max}$ for the ESA in Fig. \ref{fig1} is plotted versus its efficiency $\eta$ for five different values of the Q reduction factor $\rho$ (inverse of bandwidth-increase factor) made possible by the dispersive tuning.  The strong $\rho$ dependence of the curves can be seen in this figure.  Comparison of these curves computed directly from the bandwidth using the reflection coefficient ($|\Gamma|^2$) shows that the $\alpha_{\rm max}$ in Figs. \ref{fig2} and \ref{fig4} are accurate to a dB or so; also compare these curves with the bandwidth curve in \cite[fig. 5]{Yaghjian-IET}.
\begin{figure}[h]
\vspace{-6mm}
\hspace{-5mm}
\includegraphics[height=1.8in,width=3.7in]{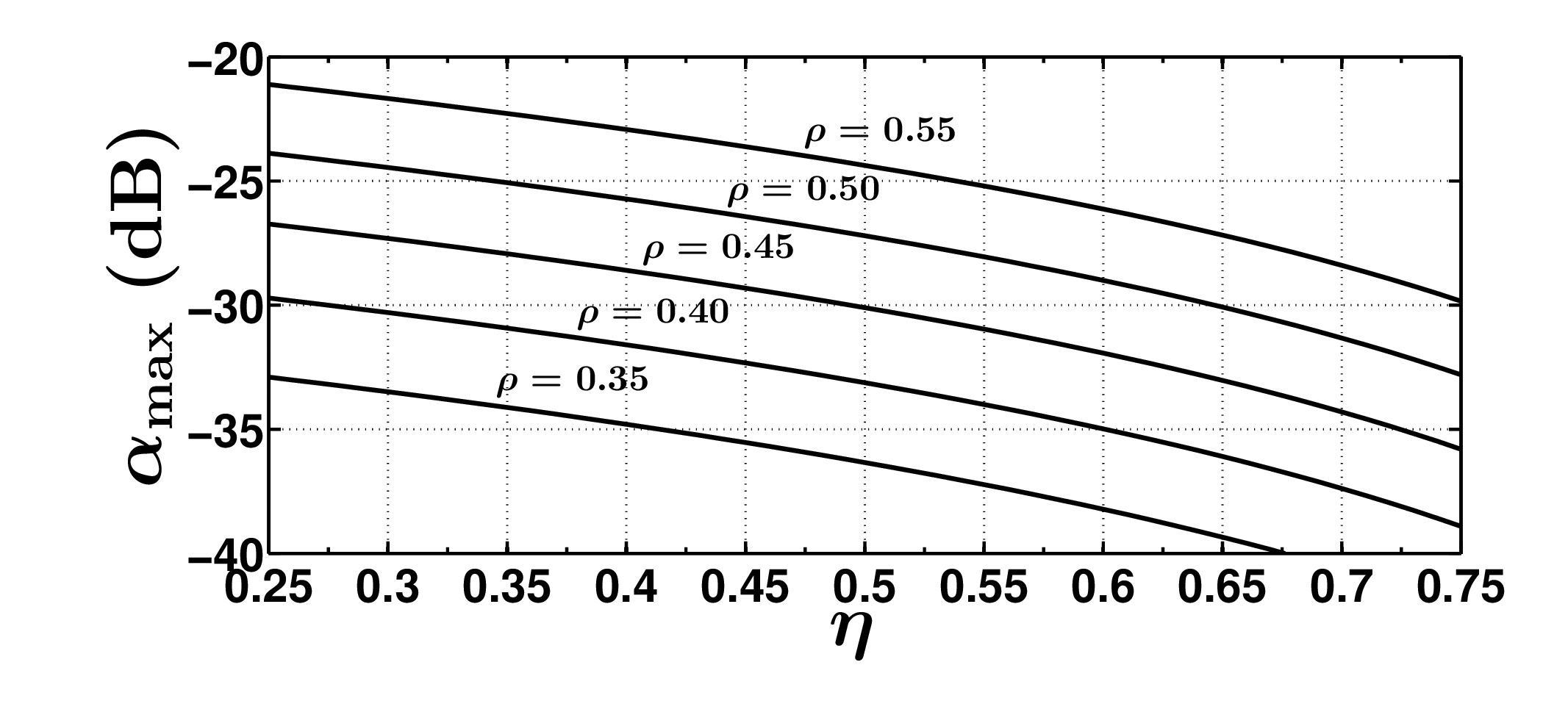}
\vspace{-7mm}
\caption{The maximum power-drop factor $\alpha_{\rm max}$ vs the efficiency $\eta$ for five different values of Q reduction factor $\rho$ (inverse of bandwidth-increase factor).}
\label{fig2}
%\mbox{}\\[-5mm]
\end{figure}
\par
In Fig. \ref{fig4}, taken from \cite{Yaghjian-Reducing}, curves for the efficiency $\eta$ versus $1/\rho$ (factor increase in fractional bandwidth) are plotted for bandwidth power drops of $\alpha_{\rm max} =  -20, -25$, and $-30$ dB.  Figs. \ref{fig2} and \ref{fig4} show that for a given efficiency the increase in bandwidth enabled by dispersive tuning becomes less with larger bandwidth drops.
\par
Typical reflection-coefficient curves versus frequency are plotted in Fig. \ref{fig5}, taken from \cite{Yaghjian-Reducing}, for dispersive and nondispersive tuning of an antenna with an original lossless quality factor of $40$, an efficiency $\eta\approx 0.5$, and a doubling of the bandwidth at $\alpha_{\rm max}\approx -27$ dB, that is, $Q_0 = Q_{0Z} = 20$ (with $\eta = 0.5$) and $Q= Q_Z= 10$ (with $\eta = 0.5$).  As the bandwidth drop falls below about $-10$ dB, the improvement in bandwidth increases monotonically from about $25\%$. These reflection-coefficient curves reveal that the dispersive tuning improves the bandwidth until the bandwidth drop equals about $-5$ dB ($\alpha_{\rm max} \approx 0.3$).  The $|\Gamma|^2$ curves in Fig. \ref{fig5} are not perfectly symmetric about the origin as in the $|\Gamma|^2$ formula (\ref{Gam}) because of the $O[(\Delta\w/\w)^3]$ term mentioned in Footnote \ref{Dcube}.
\begin{figure}[h]
\vspace{-2mm}
\hspace{-5mm}
\includegraphics[height=1.8in,width=3.7in]{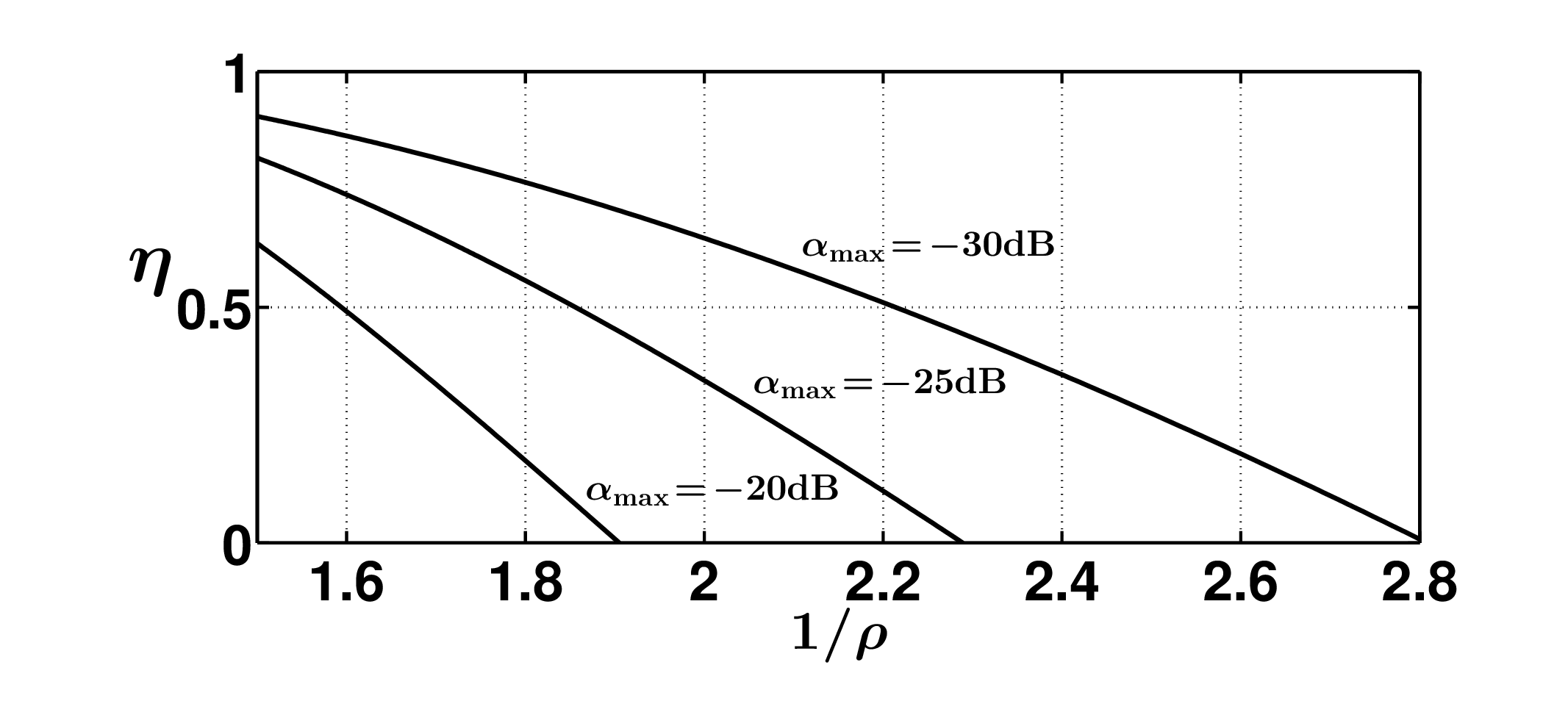}
\vspace{-9mm}
\caption{Efficiency $\eta$ versus factor increase in fractional bandwidth ($1/\rho$) for three different power-drop factors $\alpha_{\rm max}$.}
\label{fig4}
%\mbox{}\\[-6mm]
\end{figure}
\begin{figure}[h]
\vspace{-3mm}
\hspace{-1mm}
\includegraphics[height=1.7in,width=3.625in]{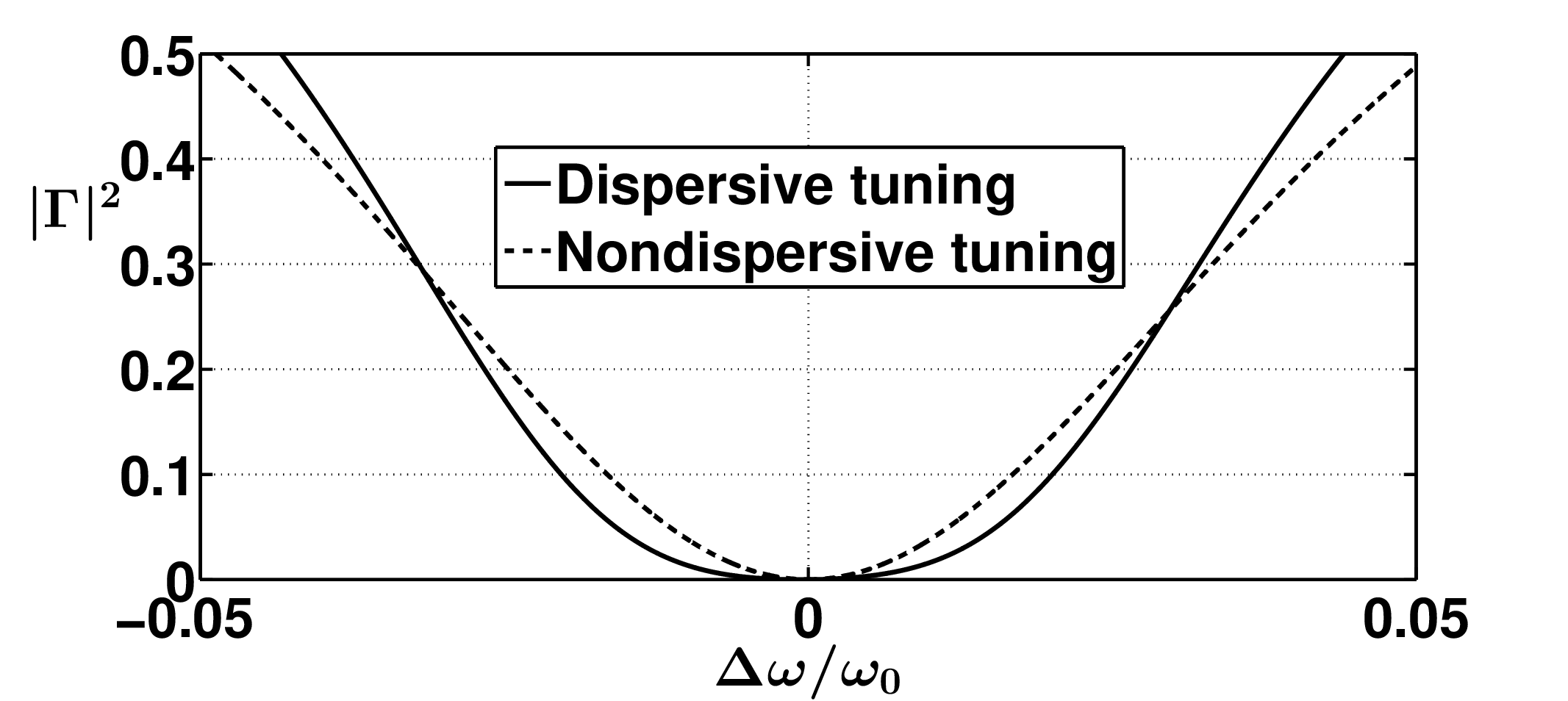}
\vspace{-7mm}
\caption{Magnitude squared of the reflection coefficients versus frequency with dispersive and nondispersive tuning for the original lossless antenna quality factor of $40$,  efficiency $\eta = 0.5$ (so $Q_0\! =\! Q_{0Z}\! =\! 20$), and a doubling of the bandwidth at $\alpha_{\rm max}\!\approx\! -27$ dB ($Q \!= \!Q_Z\! =\! 10$).  Both curves are approximately quadratics near $\Delta\w/\w_0=0$ with $|\Gamma|^2 \approx Q_Z^2(\Delta\w/\w_0)^2$; see (\ref{Gam}).}
\label{fig5}
%\mbox{}\\[-6mm]
\end{figure}
\par
As mentioned above, Al\`u has also introduced the innovation of using a simple equivalent circuit to produce the required dispersive tuning capacitance, specifically, a capacitor in parallel with a series RLC, thus eliminating the need for a capacitor filled with Lorentzian material \cite{Alu}. 
\par
In summary, using dispersive Lorentzian permittivity (or permeability) tuning, the quality factors and thus the Chu/CR lower bound for isolated resonances or antiresonances of electrically small lossy magnetic- (or electric-) dipole antennas can be halved ($\rho=0.5$), that is, the bandwidth doubled, for power-drops of the input impedance bandwidths of about $-25$ dB and radiation efficiencies of about $50\%$ ($\eta = 0.5$).  The required bandwidth power-drop decreases with increasing efficiency and/or  bandwidth improvements, so the same doubling of bandwidth can be obtained if the efficiency is raised but at a lower bandwidth power-drop.   Conversely, more than a doubling of the bandwidth can be obtained for the same bandwidth power-drop if the efficiency is lowered.  Moreover, all these increased bandwidths (decreased quality factors) occur at single isolated resonances (at the Lorentzian resonant frequency $\w_0$), as plots of $X(\w)$ versus $\w$ confirm, for example, in Fig. \ref{RX}, and thus they are not produced by the multiple resonances and antiresonances of Bode-Fano tuning, either explicitly or implicitly; see Section \ref{BFM}.  Consequently, dispersive tuning, unlike Bode-Fano tuning, does not increase the ``group delay'' of the fields radiated by the antenna. In addition, there appears to be no reason that Bode-Fano tuning with multiple resonances/antiresonances could not be combined with dispersive tuning to further increase the bandwidth.
\section{Conclusion}
A brief historical introduction begins by tracing the concept of quality factor from K.S. Johnson's work around 1914 to Wheeler's seminal 1947 and 1958 papers (in which he determines the quality factors of electrically small electric and magnetic dipoles), and ends with an outline of the later developments of the lower bounds on quality factor for spherical antennas.  We then derive useful expressions for the robust input-impedance $Q_Z(\w)$ quality factor that always gives an accurate VSWR fractional bandwidth of antennas for isolated resonances/antiresonances ($Q_Z(\w)\neq 0$) at a small enough bandwidth power drop.  For closely spaced multiple resonances/antiresonances it is shown that Bode-Fano tuning can increase the isolated-resonance/antiresonance fractional bandwidth by a maximum factor of about $4$ (and more realistically about $2$)  for bandwidth power drops between about $-4$ dB   and $-11$ dB.  After formulating and applying the method of determining the conventional and complex-energy quality factors from RLC circuit models of antennas, new field-based quality factors $Q(\w)$ are derived for antennas with known fields produced by an input current $I$.  These field-based $Q(\w)$ are remarkably robust because they are identical in value to $Q_Z(\w)$ when the input impedance is available.  Like the impedance-based $Q_Z(\w)$, the field-based $Q(\w)$ is independent of the choice of origin for the fields of the antenna and is impervious to extra lengths of transmission lines and surplus reactances.  These robust field-based quality factors are used to derive new lower bounds on the quality factors (upper bounds on the bandwidths) of spherical-mode antennas that improve upon the previous Chu/CR lower bounds for spherical modes.
\par
Harrington's maximum gain formula for a finite degree number $N$ of spherical modes is combined with the recently determined general formula for the maximum significant radius of the reactive power of an antenna to derive a criterion for antenna supergain.  Maximum gains versus minimum quality factors for self-tuned spherical antennas are determined using the improved lower bounds on quality factors and plotted for values of $ka = 0.2,\, 0.5$, and $1.0$.
\par
Lastly, reduced antenna quality factors allowed by dispersive tuning are derived and applied to show that  for an elec\-trically small electric or magnetic dipole, the traditional Chu/CR lower bounds can be overcome by about a factor of two (doubling the bandwidth) for power drops of about $-25$ dB and radiation efficiencies of about $50$\%.  Further increases in bandwidth can presumably be obtained with Bode-Fano mul\-tiple resonances/antiresonances, albeit with increased distortion of the antenna's transmitted signal caused by the Bode-Fano increased phase change (group delay) across the bandwidth.
\section*{Acknowledgments}
This paper benefited from discussions with (in alphabetical order)  A. Al\`u, S.R. Best, O. Breinbjerg, M. Gustafsson,  A.K. Iyer, S. Maci, E. Martini, J. Rahola, and R.W. Ziolkowski.  The research was supported under the U.S. Air Force Office of Scientific Research (AFOSR) Grant \# FA9550-22-1-0293 through A. Nachman.
\end{document}